\documentclass[conference]{IEEEtran}

\usepackage{amsmath,amssymb,amsfonts}
\usepackage{graphicx}
\usepackage{textcomp}
\usepackage{xcolor}
\usepackage{caption}
\usepackage{subcaption}
\usepackage[hyphens]{url}
\usepackage{caption}
\usepackage{algorithm}
\usepackage{algorithmicx}
\usepackage{algpseudocode}
\usepackage{multirow}
\usepackage{tikz}
\usepackage{color}
\usepackage{svg}
\usepackage{xspace}
\usepackage{listings}

\definecolor{datatypecolor}{rgb}{0.75, 0.0, 0.2}
\definecolor{keywordcolor}{rgb}{0.01, 0.28, 1.0}
\algnewcommand{\algorithmicforeach}{\textbf{for each}}
\algdef{SE}[FOR]{ForEach}{EndForEach}[1]
  {\algorithmicforeach\ #1\ \algorithmicdo}% \ForEach{#1}
  {\algorithmicend\ \algorithmicforeach}% \EndForEach

\usepackage{cleveref}
\crefname{section}{\S}{\S\S}

\usepackage{cite}

\newcommand{\Xlabel}{ACS}
\newcommand{\XlabelHW}{\texttt{ACS-HW}}
\newcommand{\XlabelSW}{\texttt{ACS-SW}}
\newcommand{\XlabelSWSim}{\texttt{ACS-SW-Sim}}
\newcommand{\cudagraph}{\texttt{CUDAGraph}}

\newcommand{\kernelready}{\texttt{ready}}
\newcommand{\upstream}{upstream}

\newcommand{\ready}{\texttt{ready}}
\newcommand{\pending}{\texttt{pending}}
\newcommand{\executing}{\texttt{executing}}
\newcommand{\scheduledlist}{\texttt{scheduled\_list}}

\newcommand{\schedulingwindowhw}{scheduling window}

\definecolor{darkblue}{rgb}{0.0, 0.0, 0.55}
\newcommand\insertion[1]{{\color{black}{#1}}} 
\newcommand\insertionasplos[1]{{\color{black}{#1}}}

\newcommand\insertionhpca[1]{{\color{black}{#1}}\xspace}
\newcommand\hpcarebuttal[1]{{\color{black}{#1}}\xspace}

\newcommand\deletion[1]{} 
\newcommand\iscadeletion[1]{} 

\def\resulttabmode{0}   % display results in tables
\def\outlinemode{0}     % display paper outline
\definecolor{amber}{rgb}{1.0, 0.49, 0.0}

\newcommand*\circled[1]{\tikz[baseline=(char.base)]{
            \node[shape=circle,fill,inner sep=1pt] (char) {\textcolor{white}{#1}};}}

\newcommand{\greencheck}{\multicolumn{1}{c}{\centering\textcolor{green}{\checkmark}}}
\newcommand{\redcross}{\multicolumn{1}{c}{\centering\textcolor{red}{\textbf{x}}}}

\usepackage{comment}

\if\resulttabmode0
\newcommand{\displaytable}[1]{}
\newcommand{\displaygraph}[1]{}
\else
\newcommand{\displaytable}[1]{}
\newcommand{\displaygraph}[1]{}
\fi

\if\outlinemode0
\newcommand{\outlinepoint}[1]{}
\newenvironment{outlineenv}{}{}
\newcommand{\template}[1]{}

\else
\newenvironment{outlineenv}{\begin{itemize}}{\end{itemize}}
\newcommand{\template}[1]{\textbf{\textcolor{black}{#1}}}
\newcommand{\outlinepoint}[1]{\item \template{#1}}

\fi

\usepackage[normalem]{ulem}

\begin{document}

\title{\Xlabel{}: Concurrent Kernel Execution on Irregular, Input-Dependent Computational Graphs
\vspace{-0.3cm}}

\date{}
\author{
  \IEEEauthorblockN{
    Sankeerth Durvasula\IEEEauthorrefmark{1},     
    Adrian Zhao\IEEEauthorrefmark{1},     
    Raymond Kiguru,     
    Yushi Guan,     
    Zhonghan Chen,     
    Nandita Vijaykumar\\ 
  }
}

% \author{}
\maketitle

\thispagestyle{empty}

\begin{abstract}
\deletion{Modern }GPUs are widely used to accelerate many important classes of workloads today. However, in this work, we observe that several important emerging classes of workloads, including simulation engines for deep reinforcement learning and dynamic neural networks, are unable to fully utilize the massive parallelism that GPUs offer. These applications tend to have kernels that are small in size, i.e., have few threads and thread blocks that cannot saturate the GPU's compute resources. Executing independent kernels \emph{concurrently} is a promising approach to improve parallelism and utilization. However, this inter-kernel concurrency is difficult to leverage in such workloads with existing approaches: First, the inter-kernel dependencies and computational graph are input-dependent and vary each time the application is executed. Second, the computational graphs tend to be irregular, requiring fine-grain scheduling and synchronization; thus incurring significant synchronization overheads if kernel execution is parallelized. 
In this work, we propose \Xlabel{}, a new framework that enables lightweight  detection of inter-kernel dependencies and low overhead kernel scheduling at runtime. The key idea behind \Xlabel{} is to perform inter-kernel dependency checks for a small window of kernels at runtime, similar to out-of-order instruction scheduling. This enables concurrent execution of kernels in applications whose computational graphs are input-dependent and require fine-grained scheduling. We propose \Xlabel{}-SW, a software-only open-source implementation of \Xlabel{} and \Xlabel-HW{}, a hardware-software cooperative implementation. \Xlabel-HW{} further reduces synchronization overheads by reducing communication between the CPU and GPU. We evaluate \Xlabel{} for deep RL simulation engines and dynamic and static DNNs on both real hardware and a GPU simulator. We demonstrate speedups of up to $2.19\times$ ($1.56\times$ on average) by improving GPU utilization with concurrent kernel execution.

\end{abstract}

% \vspace{-.2cm}

\section{Introduction}
Graphics Processing Units (GPUs) today are commonly used to accelerate a diverse set of applications, such as deep neural network (DNN) processing, scientific computing, graphics, and cryptography. The massive parallelism offered by GPUs enables efficient computations on large amounts of data concurrently. However, we observe that certain important classes of applications, such as simulation engines for deep reinforcement learning (RL)~\cite{brax,isaacsim, large_batch_drl, atari_gpu, samplefactory} and dynamic neural networks~\cite{ddw, d2nn, s2dnas, dynamic_nn_survey, instanas, dynamic_nn_need, dynamic_routing,  rdinet, skipconv,branchynet, blockdrop, convaig, micronet, lcnet}, are unable to fully utilize the significant compute capability that GPUs offer. This underutilization is because these applications comprise a large number of small kernels, i.e., kernels with few thread blocks that are unable to fully saturate the GPU cores. To understand the challenges in alleviating this underutilization, we evaluate two important classes of applications and introduce their properties.

\textbf{Simulation Engines for Deep RL.} With reinforcement learning (RL) an agent (for example, a robot) learns to perform tasks such as robotic locomotion, manipulation, and navigation~\cite{drive_in_a_day,inhand_reorientation} by trial and error from interactions with the environment. 
Deep RL training involves using a DNN to learn policies that optimize for rewards from data collected by interacting with a simulation environment. By leveraging the benefits of DNNs, deep RL has recently gained widespread application for many challenging and important tasks~\cite{learning_to_fly,drive_in_a_day,semantic_aware_uav_perception,interp_e2e_driving, airlearning, glide, taxim, walkminute}. Despite leveraging GPUs, a significant fraction of the deep RL runtime is the data collection phase (up to $70\%$ of the runtime), where physics simulations are used to generate training data. We observe that these physics simulations heavily underutilize the GPU, only achieving an occupancy of $34\%$ on average. The underutilization is caused by kernels that contain a small number of thread blocks that cannot fully utilize the GPU. Programming larger kernels is impractical as each instance simulates a different scenario, and large kernels would lead to thread divergence. 

\textbf{Dynamic DNNs.} Several recent types of DNNs~\cite{instanas, dynamic_routing, deepspeed_moe, ddw} have emerged as a promising approach to reduce inference latencies in resource-constrained devices by reconfiguring/specializing the architecture based on the input to the DNN. For example, InstaNAS~\cite{instanas} configures the network architecture at runtime based on the input image. Our evaluations demonstrate that, while these architectures require significantly fewer FLOPs and lower inference latencies, there is still significant underutilization of GPU resources (achieving an occupancy of only 39\% on average). Similar to the simulation engines, we find that this underutilization is caused by small kernels that do not fully utilize the GPU cores.

GPU kernels from such applications are typically executed \emph{serially}, and thus the utilization is determined by the size (i.e., the number of threads and thread blocks) of the kernel. However, we observe that many kernels are independent and thus can be executed concurrently. By concurrently executing independent kernels, we can effectively improve GPU utilization and thus performance. \insertionhpca{Existing GPU architectures allow for concurrent execution of kernels by using multiple command queues~\cite{hyperq} which are abstracted in software (such as CUDA Stream~\cite{cudastream}), allowing the programmer to identify and launch independent kernels in parallel. }
However, enabling concurrent kernel execution for these applications is still a challenging task for two major reasons.

\textbf{Challenge 1: Input-dependent computational graphs.} For these applications, the computational graph (i.e. the kernels to be executed and their dependencies) is only resolved at runtime based on the input, and each input or set of inputs leads to a different computational graph. \hpcarebuttal{This means that identifying independent kernels to launch in parallel requires performing inter-kernel dependency checks at runtime. These workloads have short running kernels that significantly exacerbate the scheduling and dependency checking overheads, making this a challenging problem to solve.} \insertionasplos{Frameworks such as CUDA Graph~\cite{cudagraph} and AMD ATMI~\cite{atmi} allow programmers to define the inter-kernel dependency information and construct a directed acyclic graph (DAG) of kernels. These frameworks enable concurrent kernel execution. However, when inter-kernel dependencies vary by input, we must incur the significant latency of constructing the dependency graph and scheduling independent kernels, every time the application is executed, significantly increasing run time (\cref{sec:motivation_cudagraph} and \cref{sec:eval}). } 

\textbf{Challenge 2: Irregular inter-kernel dependencies require fine-grain scheduling.} The computational graph for a given input tends to be highly irregular. In other words, the kernels cannot be easily partitioned into independent streams and fine-grain scheduling is required to expose inter-kernel parallelism. Thus, parallel execution of kernels requires frequent synchronization to ensure correctness, leading to significant synchronization overheads from communicating with the CPU and from kernel launches (\cref{sec:motivation_observations}).

To address these challenges, \textbf{our goal} in this work is to enable kernel concurrency with \emph{(i)} lightweight scheduling and dependency checking of kernels that can be performed at runtime and \emph{(ii)} low overhead synchronization for scheduling and kernel launch.
To this end, we propose \Xlabel{}, a new framework for \underline{A}utomatic \underline{C}oncurrent \underline{S}cheduling with two implementations: \emph{(i)} \Xlabel{}-SW, a software-only mechanism to enable lightweight kernel scheduling at runtime and \emph{(ii)} \Xlabel{}-HW: a hardware-software mechanism to further reduce synchronization overheads for efficient kernel concurrency. 

The key idea of \Xlabel{} is to perform dependency checks between sequentially launched kernels within a fixed window at runtime, similar to out-of-order instruction scheduling. We refer to this window as the \emph{scheduling window}.  When a kernel is inserted into the scheduling window, the kernels that it is dependent on are identified. As kernels complete execution, kernels in the scheduling window are marked ready based on the identified dependencies. Ready kernels can then be concurrently launched as they have no more dependencies. Since at any given time, only a small set of kernels are scheduled and tracked (instead of the entire computational graph), this approach enables efficient kernel parallelization and scheduling at runtime.
To perform dependency checks between kernels, \Xlabel{} leverages annotations from the application that specify the memory address ranges that are read/written by each kernel. This metadata is then used to identify inter-kernel dependencies at runtime when kernels are inserted into the scheduling window. Compared to prior approaches (\cref{sec:approach_prior_mechanisms}), this method alleviates the significant kernel scheduling and dependency-check overheads for kernel parallelization.  

\Xlabel{}-SW implements the above out-of-order runtime kernel scheduling in software as an application runtime system using CUDA streams. \Xlabel{}-SW however still incurs synchronization overheads from communication with the CPU and kernel launch. 
On the other hand, \Xlabel{}-HW implements the out-of-order kernel scheduler in the GPU hardware and can alleviate the synchronization overheads. We propose an efficient implementation of \Xlabel{}-HW that reduces synchronization and kernel overheads by reducing communication with the CPU. 

%  at a high level , we use a similar approach to solve this task at a high level

\insertion{Prior works such as task superscalar~\cite{task_superscalar}, carbon~\cite{carbon}, TDM~\cite{castillo_task_management} and ADM~\cite{adm} propose similar out-of-order scheduling to leverage irregular parallelism between tasks in CPU multiprocessors. However, the major challenge in CPUs is the latency of runtime dependence checking.} \insertionasplos{The primary bottleneck with GPUs is the latency for launch/signal completion of kernels rather than dependence checking (\cref{sec:depcheck}). \Xlabel{} addresses this challenge and provides an efficient approach to enable out-of-order kernel scheduling in GPUs.}

We demonstrate the effectiveness of \Xlabel{} in improving GPU utilization and thus performance for physics simulation workloads, a range of dynamic neural networks, as well as static neural networks with small kernels. We demonstrate an average speedup of up to $1.87\times$ using our software-only approach and up to $2.19\times$ from the  hardware-software implementation. The major contributions of this work are:
\begin{itemize}
    % \item We perform a characterization of
    \item We identify and characterize GPU underutilization as a result of small GPU kernels in applications with input-dependent irregular computational graphs, e.g., deepRL and dynamic DNNs.
    \item We introduce \Xlabel{}, a runtime mechanism that improves GPU utilization by enabling concurrent execution of GPU kernels with a lightweight dependency tracking and scheduling framework.
    %. , tracking irregular data dependencies between GPU kernels at runtime and concurrently execute independent kernels.
    \item We will provide an open-source software-only implementation of \Xlabel{} that can be used on real hardware to enable low overhead GPU kernel concurrency. 
    \item We evaluate the effectiveness of \Xlabel{}-SW and \Xlabel{}-HW on a range of important GPU applications and demonstrate significant speedups and improved GPU utilization.
\end{itemize}

% \section{Background}
% \input{textsrcs/background}

\section{Motivation}
% %\vspace{-0.2cm}
\subsection{Baseline GPU architecture}
\label{sec:baseline_gpu_architecture}
\begin{figure}[!htb]
\vspace{-0.1cm}
    \centering
    \includegraphics[width=0.6\linewidth]{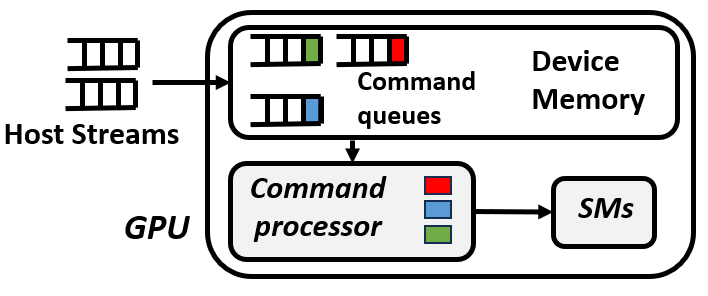}
    % %\vspace{-0.2cm}
\caption{Scheduling kernels from multiple streamsw}
\label{fig:gpuwq}
\vspace{-0.1cm}
\end{figure}

\hpcarebuttal{Figure~\ref{fig:gpuwq} shows an overview of the hardware model in modern GPU architectures~\cite{oversubscribed_queues}. The host communicates with the command processor (CP) of the GPU via a virtual memory region which is memory mapped to the GPU, accessible by the command processor. This enables communication between the CPU and GPU through entries in the command queue. The CPU transmits kernel launch packets to the GPU by writing them to the user mode command queue.  
The CP is responsible for decoding and dispatching the kernels in these command queues for execution. The CP accesses the command queue and schedules the kernels at the head for execution. This ensures that the kernels are dispatched for launch from these queues in order.
}

\label{sec:motivation}
\subsection{\textbf{Case Study 1:} Simulation Engines for Deep RL}
% %\vspace{-0.2cm}
\label{sec:motivation_brax}
Deep reinforcement learning (RL) has widely gained attention as a promising approach to learning control policies in robotics and dynamical systems for tasks such as locomotion on legged robots~\cite{glide, brax, walkminute}, dexterous hand manipulation~\cite{inhand_reorientation}, autonomous driving~\cite{interp_e2e_driving, drive_in_a_day}, and drone control~\cite{airlearning, semantic_aware_uav_perception, learning_to_fly}. Deep RL involves training a DNN to learn policies that maximize the reward, based on the actions that the agent (e.g., four-legged robot) performs in a given environment. This training process requires data from the agent interacting with a physics simulator. Typically, each training step requires data from thousands of physics simulations.
% \hpcarebuttal{The traditional approach to running several simulation engines (typically run on large, distributed CPU clusters) is computationally expensive and is a bottleneck in the training process.}
Recent works~\cite{brax, isaacsim, atari_gpu, large_batch_drl, rl_data_generation, samplefactory} accelerate this data generation phase by leveraging GPUs. GPUs can accelerate data generation by performing multiple simulations simultaneously and also parallelizing within a single simulation.
\hpcarebuttal{Hence this makes them an appropriate candidate workload for GPU execution.}
Despite GPU acceleration, the simulation/data generation phase is still the predominant computation in deep RL{\textemdash}taking about $30-70\%$ of training time depending on the complexity of the simulated environment. Thus accelerating simulation engines is critical for deep RL performance.  

\begin{figure}[!htb]
    % %\vspace{-0.3cm}
    \centering
    \includegraphics[width=.7\linewidth]{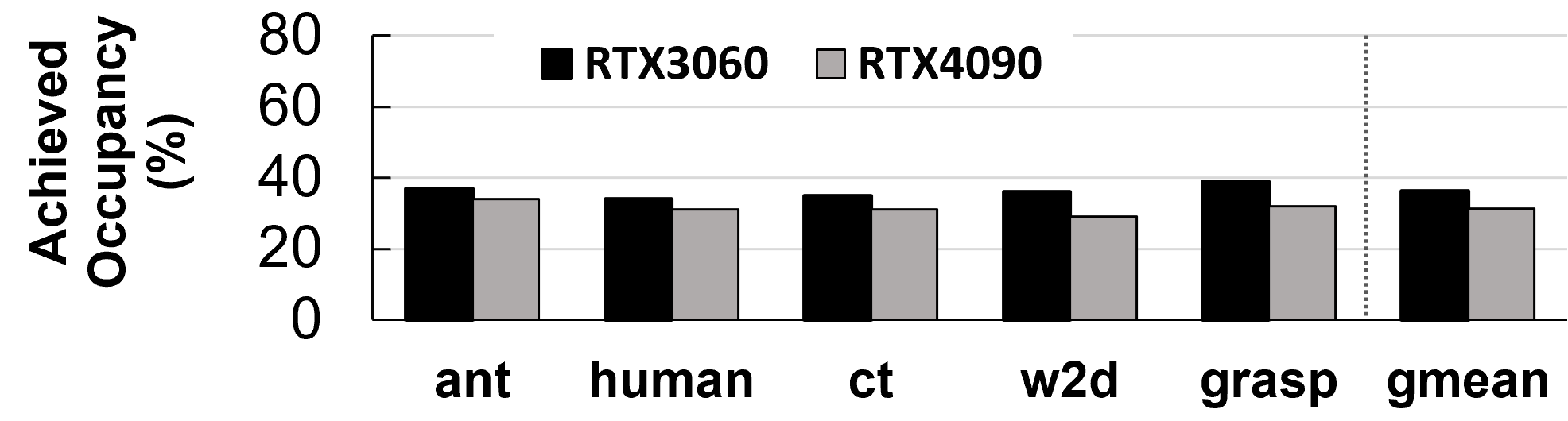}
    % %\vspace{-0.1cm}
    \caption{Simulation engines: Achieved occupancy.}
    \label{fig:utilization_brax}
    %\vspace{-0.3cm}
\end{figure}

\begin{figure}[!htb]
    \vspace{-0.cm}
    \centering
    \includegraphics[width=.7\linewidth]{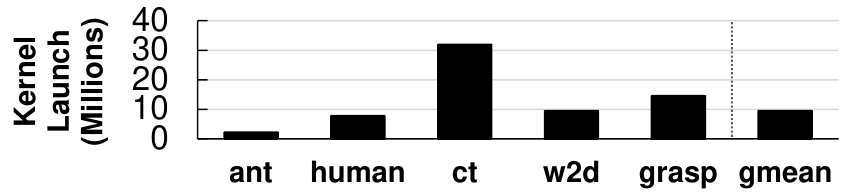}
    % %\vspace{-0.1cm}
    \caption{Simulation engines: Kernels for 1 batch of data}
    \label{brax:number_of_kernels_brax}
    \vspace{-0.1cm}
\end{figure}

\begin{figure}[!htb]
    % %\vspace{-0.3cm}
    \centering
    \includegraphics[width=.7\linewidth]{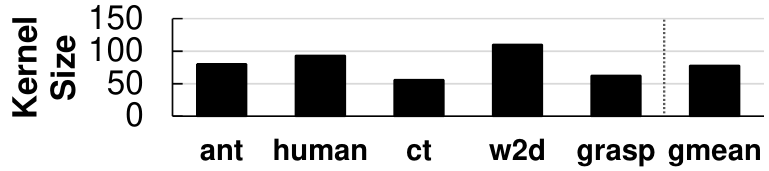}
    % %\vspace{-0.1cm}
    \caption{Simulation engines: Average kernel size (in CTAs)}
    \label{fig:avg_cta_size}
    \vspace{-0.1cm}
\end{figure}

\begin{figure}[!htb]
    % %\vspace{-0.3cm}
    \centering
    \includegraphics[width=.65\linewidth]{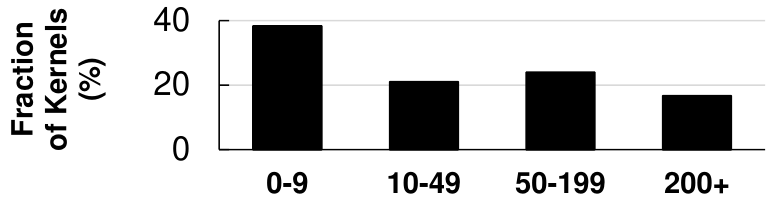}
    \vspace{-0.1cm}
    \caption{Kernel size distribution for the \texttt{ant} environment}
    \label{fig:grid_launch_size_brax}
    \vspace{-0.1cm}
\end{figure}

To evaluate the efficiency of physics simulations, we analyzed a set of physics simulations with different environments on a GPU (parameters in~\cref{sec:methodology}) with the widely used Brax~\cite{brax} framework. 
% \insertion{Brax is a simulation framework that models articulated body dynamics. Given the current state of the physical system and actions taken by an agent, its goal is to compute the next state.}
We evaluate the utilization of the GPU by measuring achieved occupancy (average ratio of active warps to the maximum supported), depicted in Fig.~\ref{fig:utilization_brax}. We find that as much as $65\%$ of the GPU cores are underutilized on average (\hpcarebuttal{on both GPUs}). To evaluate the cause of this underutilization, we analyze the number of kernel launches required to generate one batch of training data in Fig.~\ref{brax:number_of_kernels_brax}. We also present the average number of CTAs per kernel in Fig.~\ref{fig:avg_cta_size}
and depict the distribution of kernel sizes observed for the \texttt{ant} environment in Fig.~\ref{fig:grid_launch_size_brax}. We observe that physics simulations in our evaluations generate a large number of \emph{small} kernels that have few threads and CTAs. 
This is a fundamental problem because the simulation engine cannot be efficiently mapped into large kernels as the different threads will likely diverge in the execution path. 
This is because each thread typically simulates a different scenario in the environment. 
\insertion{Thus the application is instead programmed as a large number of short-running kernels.} 
% \hpcarebuttal{Each step computes the next state starting from the current state and action inputs following a dynamic execution path (which consists of checking for collisions in the neighborhood of each rigid body, impulse computation for each collision, etc.).}  As a result, each kernel cannot fully utilize all the GPU's available parallelism leading to significant underutilization.
This phenomenon has also been observed by recent works~\cite{rlscope, rl_data_generation}.% that survey simulators for deep RL. 

% %\vspace{-0.1cm}

% \vspace{-0.1cm}

\vspace{-0.1cm}

% %\vspace{-0.2cm}

%\vspace{-0.2cm}

\subsection{\textbf{Case Study 2:} DNNs with dynamic irregular graphs}
% %\vspace{-0.1cm}
\label{sec:motivation_d2nn}

\begin{outlineenv}

\outlinepoint{Efficient network architectures for edge devices and dynamic computation}
Recent research has extensively investigated specialized DNNs for edge devices with limited compute resources and power budgets as direct deployment of large neural network architectures on these devices leads to high-inference times.
Automated DNN architecture design (neural architecture search) is a promising approach to generate faster neural network architectures while retaining or improving accuracy~\cite{nasnet, amoebanet, darts, randomwire}. 
These optimized architectures tend to have irregular elaborate connections between convolution operations. 
Fig.~\ref{fig:cnn_irregular} depicts an example DNN with irregular structure.
Additionally, an emerging trend in recent research~\cite{deepspeed_moe} shows that \emph{dynamic inference models}~\cite{d2nn, ddw, s2dnas, nasood, instanas, condconv, branchynet, speechmoe, outrageously_large_moe, rdinet, skipconv, blockdrop, convaig, micronet, lcnet} are very promising to significantly reduce inference latency and FLOPs. With these dynamic inference models, the path of execution through the network is determined by the \emph{input}. Thus, the computational graph is not known ahead of time. For example, Fig.~\ref{fig:cnn_dynamic} shows an example CNN model with different paths of execution based on the input~\cite{instanas}. 

Similar to~\cref{sec:motivation_brax}, we evaluate the efficiency of these workloads on a GPU (\hpcarebuttal{an NVIDIA RTX 3060 and an NVIDIA RTX 4090}) and depict the resulting utilization in Fig.~\ref{fig:ddnn_utilization} (evaluation and workload settings are in~\cref{sec:methodology}). We find that the total achieved occupancy is around $39\%$ in the InstaNAS-A~\cite{instanas} workload \hpcarebuttal{for both GPUs}. Similar to the simulation engines, we root cause this underutilization to the existence of a large number of small kernels, as depicted in Fig.~\ref{fig:ddnn_grid_launch_size}, where a large fraction of the kernels have fewer than 200 CTAs. Thus, these small kernels are unable to fully utilize the GPU. In these workloads, the small kernels are due to convolution layers that were optimized for fewer FLOPs with smaller filters.

\begin{figure}[!htb]
    \vspace{-0.3cm}
    \centering
% \begin{subfigure}[t]{0.235\textwidth}
\begin{subfigure}[t]{0.28\textwidth}
    \centering
    \includegraphics[width=\linewidth]{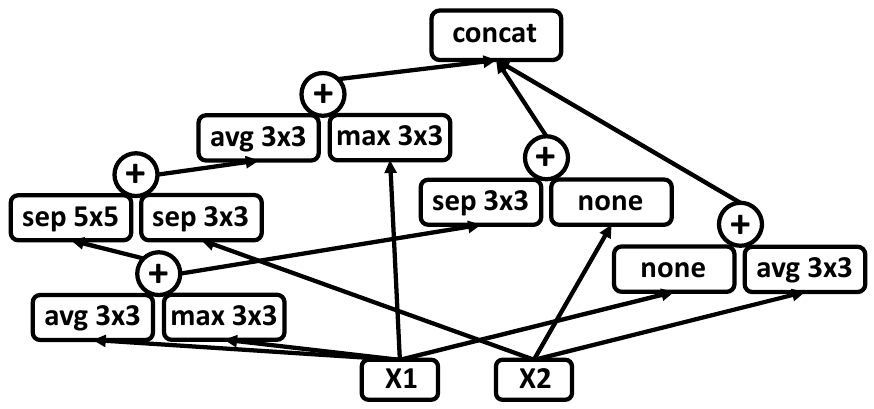}
    % \vspace{-0.2cm}
    \caption{Amoebanet~\cite{amoebanet}}
    \label{fig:cnn_irregular}
\end{subfigure}~
\begin{subfigure}[t]{0.2\textwidth}
% \begin{subfigure}[t]{0.235\textwidth}
    % \vspace{-0.1cm}
    \centering
    \includegraphics[width=\linewidth]{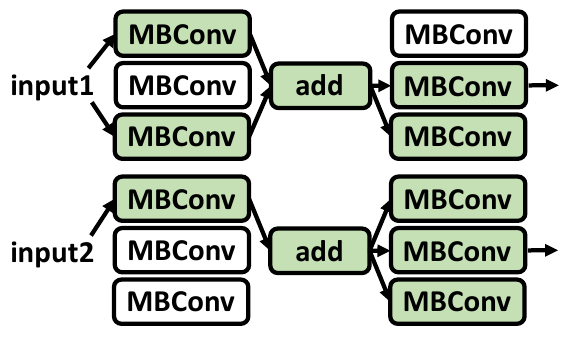}
    % \vspace{-0.2cm}
    \caption{InstaNAS~\cite{instanas} with MBConv~\cite{mobilenet} units}
    \label{fig:cnn_dynamic}
\end{subfigure}
% \vspace{-0.2cm}
\caption{DNNs with irregular or dynamic structures}
\label{fig:cnn_structures}
% \vspace{-0.3cm}
\end{figure}

\begin{figure}[!htb]

    % \vspace{-0.3cm}
    \centering
    \includegraphics[width=.7\linewidth]{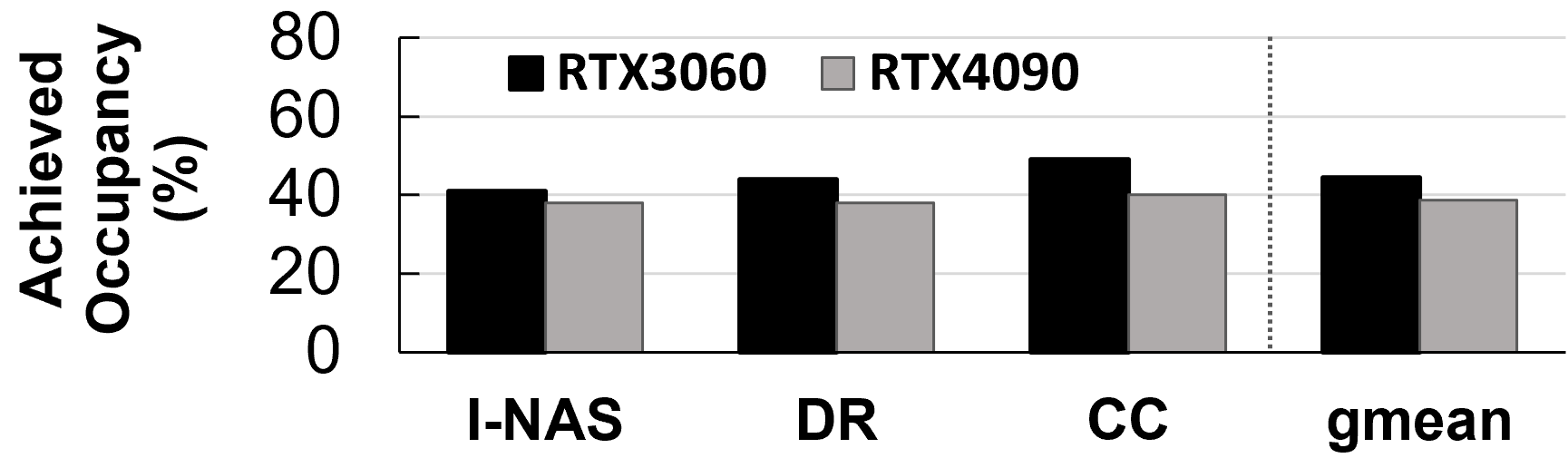}
    % \vspace{-0.2cm}
    \caption{Dynamic neural networks: Achieved occupancy}
    \label{fig:ddnn_utilization}
    % \vspace{-0.3cm}
\end{figure}

\vspace{-0.1cm}
\begin{figure}[!htb]
    % %\vspace{-0.2cm}
    \centering
    \includegraphics[width=.75\linewidth]{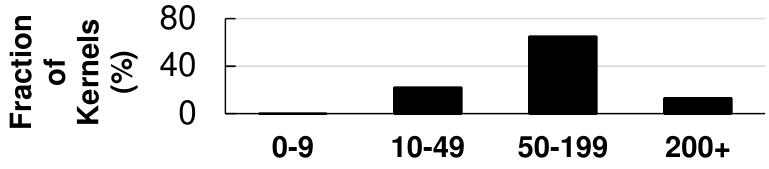}
    % \vspace{-0.2cm}
    \caption{Kernel size distribution (in CTAs) for InstaNAS-A~\cite{instanas}}
    \label{fig:ddnn_grid_launch_size}
    % \vspace{-0.5cm}
\end{figure}

\end{outlineenv}

\vspace{-0.1cm}

\subsection{Key Observations}
%\vspace{-0.1cm}
\label{sec:motivation_observations}
\label{sec:motivation_cudagraph}

\begin{outlineenv}
While small-sized kernels lead to underutilization, we observe that there are typically many kernels that can be executed \emph{concurrently}. Thus we can improve GPU utilization and reduce runtimes by identifying independent kernels and scheduling them for concurrent execution. However, this is a challenging task for these classes of applications for the following reasons. 

\textbf{(1) Input-dependent kernel dependencies.} The computational graph, and hence, the dependencies between kernels are only determined at \emph{runtime} for each input. For example, with the instance-aware dynamic DNNs~\cite{instanas, s2dnas,d2nn, ddw} described in~\cref{sec:motivation_d2nn}, for the classification inference task, the computational graph is different for each image. \insertionasplos{As a result, the determination of kernel dependencies and scheduling of kernels for the entire computational graph needs to be done for \emph{each input}. This adds significant latencies to the runtime.   

CUDA Graphs~\cite{cudagraph} and AMD ATMI~\cite{atmi} are software frameworks that allow developers to specify dependencies between different kernels as edges of a directed acyclic graph (DAG). The challenge with this approach is that the DAG needs to be constructed in full (with dependencies, kernel launches, and barriers determined) before the application is executed on the GPU, \emph{for each input}.  
%the entire graph must be constructed in advance, and execution can only begin once the entire DAG information has been entirely transferred. 
This process adds high latency in compiling the complete dependency information. We perform an experiment to measure the DAG construction and launch time on Brax~\cite{brax} simulation engine (\cref{sec:methodology}) compared to the program execution time, shown in Fig.~\ref{fig:cudagraph_constructiontime}. We observe that the time taken to construct the graph is exceedingly high (average of $47\%$ of overall execution time).}

\begin{figure}[!htb]
    \vspace{-0.3cm}
    \centering
    \includegraphics[width=0.7\linewidth]{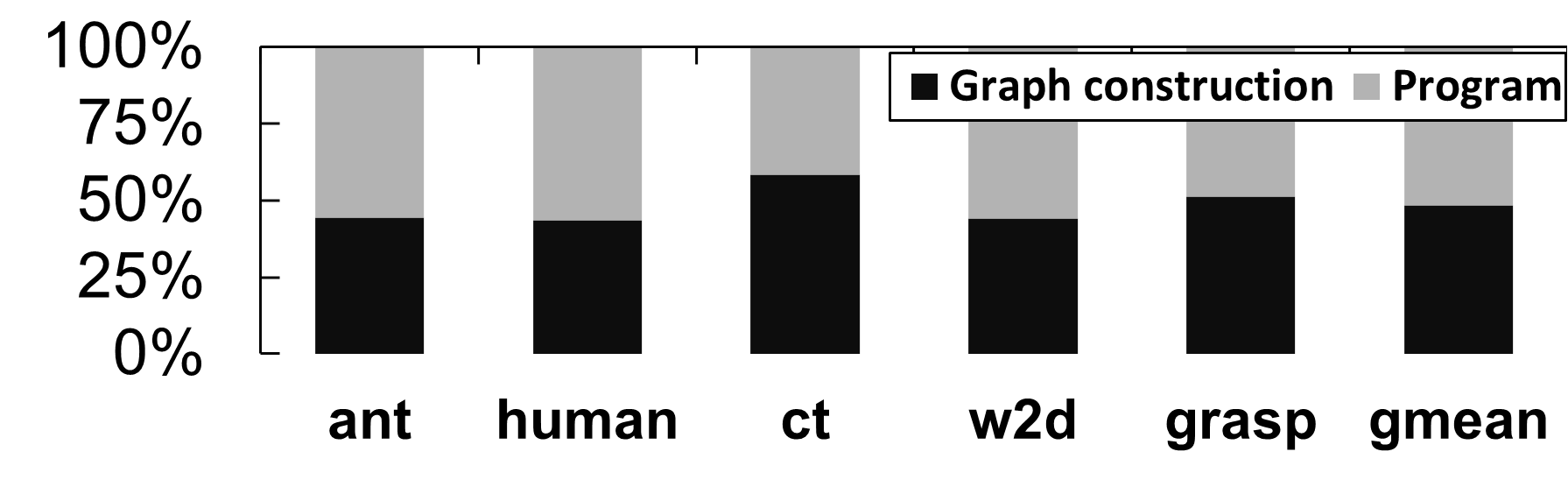}
    \vspace{-0.2cm}
    \caption{\insertionasplos{DAG construction time as \% of execution time}}
    \label{fig:cudagraph_constructiontime}
    \vspace{-0.3cm}
\end{figure} 

% \insertionhpca{One way to utilize a directed acyclic graph (DAG) approach without the need to reconstruct the graph repeatedly is by creating a super-network. A super-network is a DAG that encompasses a comprehensive set of all potential dependencies between launchable kernels. To achieve this, the super-network is launched with a CUDA graph, wherein each kernel has an extra argument flag that denotes whether the kernel should execute or not. However, this is only possible to do when all execution paths are known beforehand. }

Similarly, recent works for DNNs~\cite{ios, nimble, ooobackprop} perform kernel scheduling, fusion, or parallelization for better GPU utilization. These works, for example, partition the computational graph into independent sub-graphs that are scheduled into multiple streams. However, this scheduling and partitioning is too time-consuming to be done for each input at runtime and thus cannot be applied to these classes of workloads.

\textbf{(2) Irregular kernel dependencies.}
These classes of applications have \emph{irregular} computational graphs that are challenging to easily partition into CUDA streams (\cref{sec:motivation_d2nn}). Popular deep learning frameworks~\cite{tensorflow,pytorch} use a single stream by default. The stream abstraction works best if the entire graph can be partitioned into independent streams of kernels. However, these graphs with irregular dependencies would require fine-grained scheduling and heavy use of synchronization (e.g., cudaDeviceSynchronize and cudaStreamSynchronize) when parallelizing using CUDA streams. This synchronization may lead to large overheads as it requires communication between the GPU and CPU.
Fig.~\ref{fig:synchronization_overheads} depicts the different overheads when CUDA streams are used for fine-grained scheduling with irregular graphs: kernel launch overheads~\circled{1}, CPU execution overheads~\circled{2} and the synchronization overheads~\circled{3}. Based on our profiling, the synchronization and launch overheads vary between $5$-$20us$.
\insertionasplos{CUDA Graphs~\cite{cudagraph} and ATMI~\cite{atmi} can eliminate the synchronization and kernel launch overhead. However, for input-dependent graphs, as demonstrated in (1), this benefit is lost due to DAG construction overheads.}

    \vspace{-0.1cm}
\begin{figure}[!htb]
    % %\vspace{-0.3cm}
\centering
\includegraphics[width=0.65\linewidth]{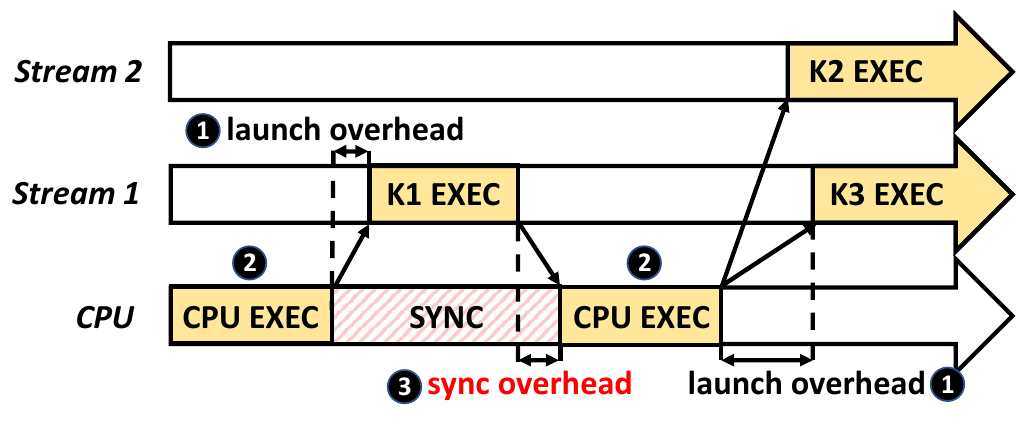}
    % \vspace{-0.2cm}
\caption{Kernel launch and synchronization overheads} 
\label{fig:synchronization_overheads}

    \vspace{-0.cm}
\end{figure}
    \vspace{-0.5cm}

\end{outlineenv}

\section{Approach}
% %\vspace{-0.2cm}
Our \textbf{goal} in this work is to design a framework that enables efficient concurrent execution of GPU kernels \emph{(i)} whose computational graph may only be known at runtime, \emph{(ii)} without incurring significant synchronization overheads. To this end, we introduce \Xlabel{}, a new framework that concurrently schedules independent kernels with a lightweight runtime mechanism. 

\subsection{Prior Mechanisms}
\label{sec:approach_prior_mechanisms}
\hpcarebuttal{We consider the baseline GPU architecture as described in~\cref{sec:baseline_gpu_architecture}.} The GPU runtime can launch kernels into different streams. These streams are mapped to one of the command queues in the device-mapped memory of the GPU. The command processor schedules kernels at the head of these queues concurrently, thus enabling concurrent kernel execution. 
However, neither the command processor nor the kernel launch packets in the command queues have information on inter-kernel data dependencies. Kernels in different queues are assumed to be independent of each other and all kernels within the same queue are executed in order. Hence, in order to leverage parallelism in kernel executions, the task of checking inter-kernel dependencies and determining the kernels which can execute concurrently (and thus scheduling into different queues) \emph{has to be done by the host application.} However, this is a problem, as this adds significant dependency-checking/scheduling latency to the run time. It also requires communication with the host (through a synchronization routine) to be performed each time a kernel completes execution, adding to the overhead.  
Several prior works describe approaches to efficiently schedule kernels into multiple streams. Fig.~\ref{fig:inorderooo} depicts approaches to scheduling a computational graph (Fig.~\ref{fig:dependency_graph}). Fig.~\ref{fig:inorder} is the baseline approach used by many existing frameworks~\cite{tensorflow, pytorch}, where a single CUDA stream is used to execute all kernels serially. This approach leads to underutilization (\cref{sec:motivation_d2nn}). Fig.~\ref{fig:barriersync} shows prior works~\cite{ios, nimble} that use the computational graph to identify independent kernels and the \emph{entire graph} is scheduled ahead of time into multiple CUDA streams. However, this fine-grained scheduling and synchronization leads to large overheads.

\begin{figure}[!htb]
    \centering
\begin{subfigure}{0.235\textwidth}
    \centering
    \includegraphics[width=.75\linewidth]    {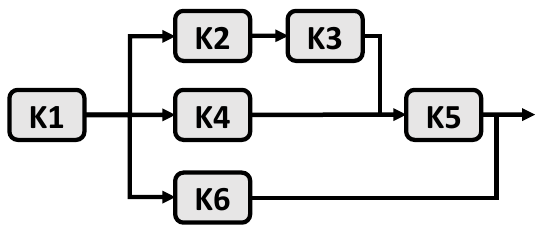}
    \vspace{-0.2cm}
    \caption{Kernel dependencies}
    \label{fig:dependency_graph}
\end{subfigure}
~\begin{subfigure}{0.235\textwidth}
    \centering
    \includegraphics[width=.8\linewidth]{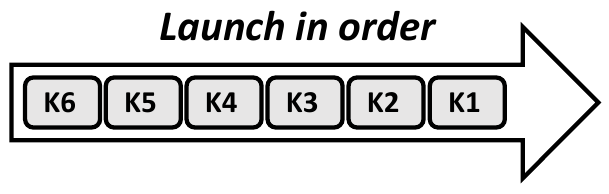}
    \vspace{-0.2cm}
    \caption{Single stream execution}
    \label{fig:inorder}
    \end{subfigure}
\begin{subfigure}[t]{0.49\textwidth}
    \centering
    \includegraphics[width=0.5\linewidth, height=0.23\linewidth]{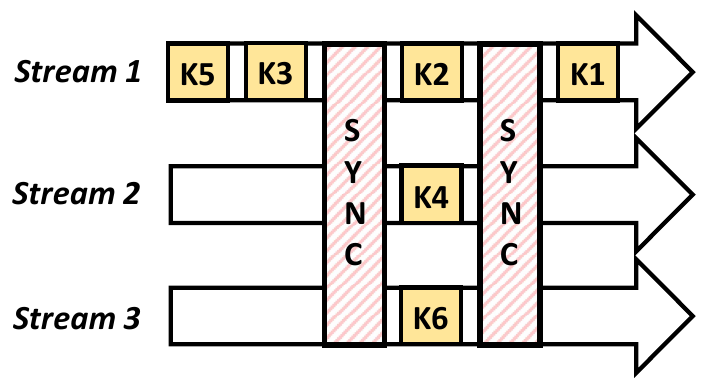}
    \vspace{-0.2cm}
    \caption{Multiple streams with synchronization between streams}
    \label{fig:barriersync}
\end{subfigure}
\caption{Scheduling kernels in a computational graph}
\label{fig:inorderooo}
\vspace{-0.3cm}
\end{figure}

\insertionhpca{One way to avoid using a device-level synchronization (like cudaDeviceSynchronize) and enable asynchronous execution of kernels without communication with the CPU is to use events provided by the CUDA stream management API. Events serve as signaling mechanisms to indicate the occurrence of specific operations in a stream. This allows synchronization between kernels across streams through the cudaStreamWaitEvent API, facilitating asynchronous kernel execution without blocking the host.
By strategically placing events and using cudaStreamWaitEvent, it is possible to orchestrate the order in which kernels are executed on the GPU without communication with host. However, this approach still requires deriving dependencies between all kernels beforehand, and thus incurs significant scheduling overhead.}

Another set of approaches~\cite{ooobackprop, mxnet, nimble}, define static dependencies between kernels as a DAG, which is then scheduled with DAG frameworks (CUDA Graph~\cite{cudagraph}/ATMI~\cite{atmi}). These approaches cannot be applied to input-dependent computation graphs, as constructing the entire computational graph is too time-consuming to be done at runtime. To convey the DAG information, ATMI sends barrier packets~\cite{hsa_barrier} along with kernel launch packets to the command queue. A barrier packet~\cite{hsa} is a 64-byte data \insertionhpca{packet that contains id information about a kernel and a set of kernels that depend on it. This packet can be inserted into the command queue by the device runtime.} The barrier packet blocks the launching of dependent kernels until the independent kernel completes execution. 
The barrier packet however does not contain any information regarding the current status of the executing kernels in the GPU and thus cannot perform any additional runtime reordering of kernels. It simply follows the dependencies already specified by the DAG. \insertionhpca{While it is possible to devise a framework that dynamically launches barrier packets and launch commands onto the GPU command queue in memory, this would require hardware support and would still incur synchronization overheads with the CPU. Our approach is specifically designed to mitigate this scheduling cost by avoiding direct communication from the GPU to the CPU, thereby reducing potential overheads.}
% This is because such an approach would necessitate adding hardware support to signal the completion of kernel execution to the host-side command queue. 

%and the incurred synchronization overheads are significant. 
Persistent threads (PT) eliminate the scheduling and launch overheads but are only effective when all kernels are homogeneous~\cite{atos} \iscadeletion{(\cref{sec:persistent_threads})}. CUDA dynamic parallelism~\cite{cdp} (CDP) \insertionhpca{or AMD's device enqueue~\cite{device_enqueue} (DE)}  enables parent kernels to launch child kernels, only allowing data dependencies between one parent and its children. These workloads however involve kernels that depend on multiple kernels, and it is an open problem how to use CDP for these types of dependencies.

We summarize different approaches for parallel kernel scheduling in Table~\ref{table:comparison_to_works_descr}, in terms of applicability (whether input-dependent irregular workloads can be effectively mapped), synchronization/launch overheads and preparation overhead (resolving dependencies, constructing, and scheduling the computational graph). 
% %\vspace{-0.2cm}
\begin{table}[!htb]
\setlength{\tabcolsep}{0.1cm}
\centering
\footnotesize
\begin{tabular}{llll}
\textbf{Method} & \textbf{Applicability} & \textbf{Sync+Launch} & \textbf{Preparation} \\
                & \textbf{} & \textbf{Overhead} & \textbf{Overhead} \\
\hline
Multi-Stream~\cite{ios, nimble} & \greencheck{} & \redcross{} & \greencheck{} \\
\hline
DAG~\cite{cudagraph,atmi,ooobackprop} & \greencheck{} & \greencheck{} & \redcross{}\\
\hline
 PT~\cite{atos,juggler,whippletree}  & \redcross{}  & \greencheck{} & \greencheck{} \\
\hline
 CDP~\cite{cdp} / DE~\cite{device_enqueue} & \redcross{} & \redcross{} & \greencheck{}\\
\hline
 \hpcarebuttal{\Xlabel{}-SW (Our approach)}            & \greencheck{}  & \redcross{} & \greencheck{}\\
\hline
 \textbf{\Xlabel{}-HW (Our approach)}            & \greencheck{}  & \greencheck{} & \greencheck{}\\
\hline
\end{tabular}
    % %\vspace{-0.1cm}
\caption{Comparison of \Xlabel{} to other scheduling frameworks}
\label{table:comparison_to_works_descr}
    %\vspace{-0.2cm}
\end{table}
    \vspace{-0.4cm}

\deletion{Fig.~\ref{fig:inorderooo} depicts prior approaches to scheduling a computational graph (Fig.~\ref{fig:dependency_graph}). Fig.~\ref{fig:inorder} is the baseline approach used by many existing frameworks~\cite{tensorflow, pytorch}, where a single CUDA stream is used to execute all kernels serially. This approach leads to the underutilization demonstrated in~\cref{sec:motivation_d2nn}.  
Existing works~\cite{ios, nimble, ooobackprop, mxnet} use the computational graph to identify independent kernels and the \emph{entire graph} is scheduled ahead of time into multiple CUDA streams, or use cuda Graphs~\cite{cudagraph} to define static dependencies between kernels. This approach cannot be applied to input-dependent computation, as scheduling the entire computational graph is too time-consuming to be done at runtime and the incurred synchronization overheads are significant.
\begin{figure}[!htb]
    \centering
\begin{subfigure}{0.23\textwidth}
    \centering
    \includegraphics[width=.8\linewidth]{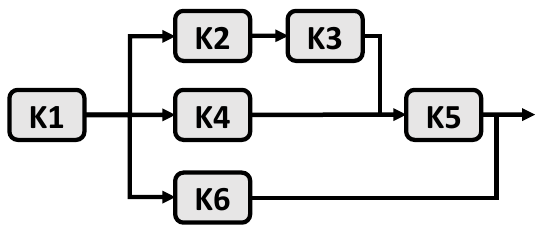}
    \caption{Dependencies between kernels}
    \label{fig:dependency_graph}
\end{subfigure}
~\begin{subfigure}{0.23\textwidth}
    \centering
    \includegraphics[width=.8\linewidth]{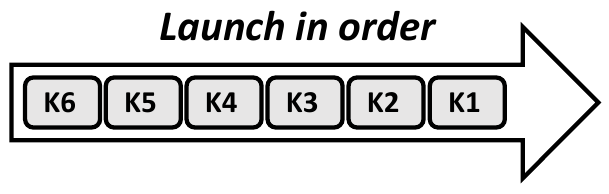}
    \caption{Single stream: Executes sequentially}
    \label{fig:inorder}
    \end{subfigure}
\deletion{
\begin{subfigure}[t]{0.49\textwidth}
    \centering
    \includegraphics[width=0.55\linewidth, height=0.3\linewidth]{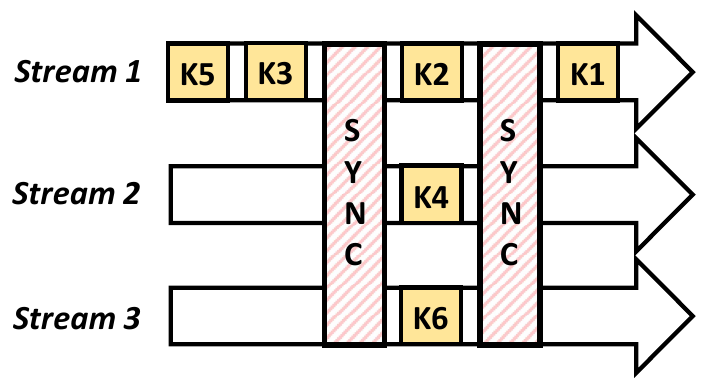}
    \caption{Multiple streams with synchronization between streams \insertionasplos{(diagrams not depicting novelty)}}
    \label{fig:barriersync}
\end{subfigure}
}
    % %\vspace{-0.1cm}
\caption{Scheduling kernels in a computational graph}
\label{fig:inorderooo}
% %\vspace{-0.cm}
\end{figure}
}

\subsection{Key Idea of \Xlabel{} }

\begin{figure}[!htb]
    %\vspace{-0.3cm}
    \centering
    \begin{subfigure}{0.24\textwidth}
        \centering
        \includegraphics[width=1.1\linewidth]{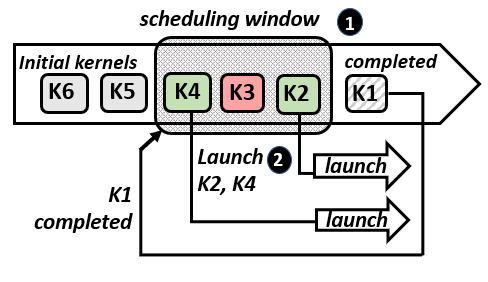}
    % %\vspace{-0.2cm}
        \caption{Out-of-order kernel dispatch from the scheduling window}
        \label{fig:outoforder}
    \end{subfigure}
    ~\begin{subfigure}{0.23\textwidth}
    \centering
    \includegraphics[width=0.75\linewidth]{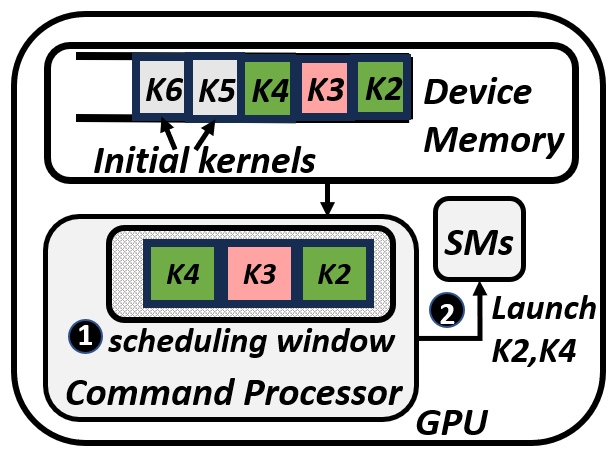}
    % %\vspace{-0.2cm}
    \caption{CP scheduling kernels in out of order manner}
    \label{fig:ooo_gpu}
    \end{subfigure}
    
    % %\vspace{-0.2cm}
    \caption{\Xlabel{}: Runtime out-of-order kernel scheduling}
    %\vspace{-0.2cm}
\end{figure}

With \Xlabel{}, the key idea is to instead perform the dependence checking and scheduling within a small window of kernels at \emph{runtime} similar to out-of-order instruction scheduling. \insertionhpca{We perform this scheduling over a single command queue (or a single initialized stream). Fig.~\ref{fig:outoforder} depicts out-of-order kernel dispatch with \Xlabel{}. Fig.~\ref{fig:ooo_gpu} shows the corresponding high-level  hardware modifications for \Xlabel{}.} A fixed number of kernels in the original stream (scheduling window~\circled{1}) are evaluated for dependencies. When a kernel completes execution, we evaluate which kernels within the scheduling window are now ready for execution~\circled{2}. 
All such kernels are marked ready and can be scheduled concurrently. 
    %\vspace{-0.1cm}

% \todo{Explain how the dependence check is handled}
We propose two implementations of \Xlabel{}: \Xlabel{}-SW, a SW-only approach and \Xlabel{}-HW, a hardware-software cooperative mechanism, which we describe in the following sections. \Xlabel{}-SW \insertionhpca{emulates the out-of-order kernel scheduling mechanism by scheduling independent kernels into multiple streams and} can be implemented with purely software changes, however the hardware support in \Xlabel{}-HW is more efficient as it also alleviates synchronization overheads. 

\subsection{Design Overview}
\label{sec:apprach_design}
% %\vspace{-0.2cm}
To design \Xlabel{} to perform the runtime kernel scheduling as depicted in Fig.~\ref{fig:outoforder}, we need \emph{(i)} a mechanism to determine inter-kernel dependencies in the scheduling window; \emph{(ii)} to identify kernels that are ready for execution; and \emph{(iii)} alleviate synchronization and kernel launch overheads.  

% There are three requirements to implement \Xlabel{}. 
%We make the following design choices in the implementation of \Xlabel{}.

\begin{figure}[!htb]%{0.5\textwidth}
    \vspace{-0.2cm}
    \centering
    \includegraphics[width=.65\linewidth]{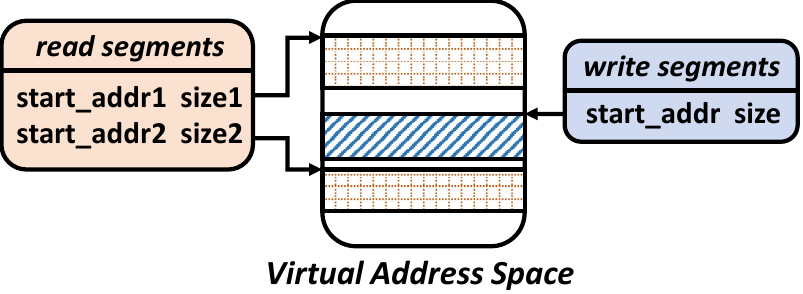}
    % %\vspace{-0.2cm}
    \caption{Memory regions written to/accessed by the kernel}
    \label{fig:memory_description}
    \vspace{-0.2cm}
\end{figure}

\textbf{Determining inter-kernel dependencies.} 
In order to determine dependencies between kernels, the application adds additional metadata to each kernel invocation. This metadata defines the range of global memory addresses that are written to and read from by each kernel. 
%\insertion{\tofix{These} are computed by a function just before the kernel launch. 
%These structures contain a list of address ranges represented with a start address and an offset (indicating the range of virtual addresses between \texttt{start\_addr} to \texttt{start\_addr+offset}). Fig.~\ref{fig:memory_description} shows the information stored in the \texttt{read\_segments} and \texttt{write\_segments}. }
This metadata is provided to \Xlabel{} by using a kernel wrapper (described in~\cref{sec:swdesign}) and can be defined by the programmer, library-writer, or compilation tools. By checking for overlaps between read segments and write segments, we determine dependencies between kernels. The kernel wrapper defines the pointers to the read and write data segments (\texttt{start\_addr}) along with the size of the segments (Fig.~\ref{fig:memory_description}). The actual virtual addresses associated with the pointers are resolved just before kernel launch in order to perform the dependence checks (\cref{sec:kernelwrappers}). We refer to these memory ranges as \texttt{read\_segments} and \texttt{write\_segments}.

\begin{figure}[!htb]
    \vspace{-0.3cm}
    \centering
    \includegraphics[width=.73\linewidth]{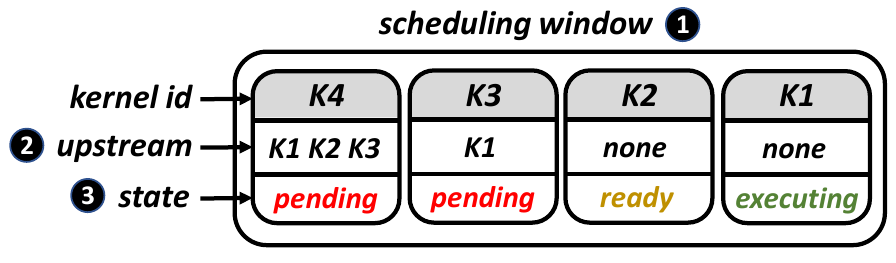}
    % %\vspace{-0.23cm}
    \caption{Kernels in the scheduling window with their state and corresponding upstream kernels (i.e., dependencies)}
    \label{fig:upstream}
    % %\vspace{-0.2cm}
\end{figure}

\textbf{Tracking kernel state at runtime.} Fig.~\ref{fig:upstream} depicts the scheduling window (\circled{1}), with the additional state required for scheduling. The kernels in the window can be \ready{}, \pending{}, or \executing{} (\circled{3}). 
Kernels in the scheduling window become ready for launch (\kernelready{}) when the kernels it is dependent on (referred to as \emph{\upstream{}} kernels \circled{2}) complete execution. For each kernel in the scheduling window, we track a list of the corresponding \upstream{} kernels. The \upstream{} kernels are determined using the above dependency checks when inserting into the scheduling window. When the \upstream{} list is empty, the kernel is marked \ready{} for execution. After each kernel completes execution, the \upstream{} list is updated for all kernels in the scheduling window. For \Xlabel{}-SW, these checks are performed in the software runtime system (\cref{sec:swdesign}), and for \Xlabel{}-HW, we implement them in hardware (\cref{sec:hwdesign}). 

    % %\vspace{-0.1cm}
    %\vspace{-0.1cm}

% Second, we need to be able to track the state of execution of GPU kernels, in order to update the \texttt{awaiting\_kernels\_list} of each kernel in the scheduling window.
% For our software-only technique, we initialize a configurable fixed number of streams at the start of the application.
% We implement a work-stealing algorithm on the ready-to-launch kernels where corresponding to each stream, we also initialize a worker thread which polls for a ready to launch kernel from the scheduling window.
% Upon receiving a ready-to-launch kernel, it launches the kernel and waits for completion with a stream level synchronization routine.
% When completed execution, this worker thread continues polling for ready-to-launch kernels and informs the mechanism to update the \texttt{awaiting\_kernel\_list} for kernels in the scheduling window.
% This ensures that no blocking operations are inserted in the application code.
% For our SW-HW solution, we implicit synchronization routine which indicated kernel completion to our mechanism.
\textbf{Eliminating CPU synchronization overheads.}
In order to eliminate synchronization and kernel launch overheads resulting from communication between the CPU and GPU, we implement the scheduling window in the GPU hardware in \Xlabel{}-HW. We design an efficient implementation of \Xlabel{}-HW that reduces communication with the CPU. The management of the scheduling window is done entirely in hardware, including the determination of \ready{} kernels. Similarly, once a kernel completes execution, the scheduling window is updated without requiring synchronization with the CPU. \deletion{We present a detailed design in~\cref{sec:hwdesign}. }
%\insertion{
%Based on the read/write segments for all kernels launched before it up to the length of the \schedulingwindow{}, the CPU software runtime computes the \upstream{} information.
%This kernel along with with the \upstream{} is passed to the GPU for execution.
%\Xlabel{}-HW implements a hardware queue to maintain information on the \schedulingwindow{} with which our mechanism can derive \ready{} kernels.
%We use a hardware signal \kernelcompletehw{} to indicate completion of kernel in the GPU. With this signal, we can update the \schedulingwindowhw{} in hardware without requiring communication with the CPU.}
% Our implementation tracks and maintains the FIFO in a circular buffer in hardware. We use an implicit synchronization routine in hardware to notify the end of execution of each kernel.

% %\vspace{-0.2cm}
\subsection{Mechanism Walkthrough}
\begin{figure}[htb!]
    \vspace{-0.2cm}
    \centering
    \includegraphics[width=.75\linewidth]{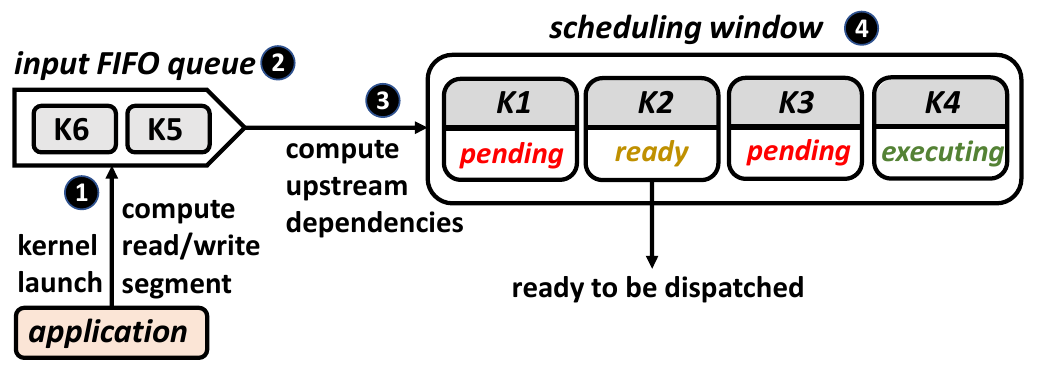}
    \vspace{-0.1cm}
    \caption{High level overview of \Xlabel{}}
    \label{fig:overview}
    \vspace{-0.3cm}
\end{figure}

% %\vspace{-0.2cm}
Fig.~\ref{fig:overview} depicts a high level walkthrough of \Xlabel{}. 
For each GPU kernel invoked by the application \circled{1}, the read and write segments are resolved (detailed in~\cref{sec:kernelwrappers}). 
All invoked kernels along with the corresponding read/write segments are entered into the input FIFO queue to await scheduling~\circled{2}. Kernels are then added to the fixed size scheduling window in a FIFO manner~\circled{3}. 
When the kernel enters the scheduling window~\circled{4}, the write segments of the current kernel are compared against read and write segments of all kernels in the scheduling window. The kernels with overlap are added to the corresponding \upstream{} kernel list and are marked \pending{}. When an \executing{} kernel completes execution, all corresponding \upstream{} kernel lists are updated. Any kernel that has an empty list is marked \ready{} for the scheduler to launch.

\section{Detailed Design}
% %\vspace{-0.2cm}
\subsection{\Xlabel{} Kernel Wrappers}
% %\vspace{-0.2cm}
\label{sec:kernelwrappers}
In order to perform runtime dependency checks, the application defines the read/write segments for each kernel. These segments are defined using a kernel wrapper, \texttt{\Xlabel{}\_wrapper} (defined in Fig.~\ref{fig:kernel_definition}). Since virtual addresses can only be resolved at runtime, the programmer instead defines a function  \texttt{get\_addresses} which populates the \texttt{\_\_read\_segments\_\_} and \texttt{\_\_write\_segments\_\_} lists (lines 6 and 7 in Fig.~\ref{fig:kernel_definition}). The \texttt{get\_addresses} function takes the kernel's launch arguments as the input arguments (lines 12 to 15). These arguments are then used to compute the read/write segments. 

Just before kernel launch, the CUDA runtime calls the \texttt{get\_addresses} function. At this point, the \texttt{\_\_read\_segments\_\_} and \texttt{\_\_write\_segments\_\_} lists are populated with the resolved virtual addresses. In our implementation of \Xlabel{}-SW, since the CUDA drivers are closed-source, we implement an intermediate user-level kernel launch function that calls the \texttt{get\_addresses} function instead. Fig.~\ref{fig:getaddress} depicts an example implementation of the \texttt{get\_addresses} function. 
\Xlabel{} assumes that the programmer or the kernel library provider has knowledge of the memory regions accessed by the kernel from the kernel function prototype. 
For a wide range of commonly used kernels, such as matrix multiplication, convolution, addition, etc., which operate on data stored as contiguous regions in memory, this task is straightforward. 
Additionally, the \texttt{get\_address} function can be obtained using a static binary analysis tool like GPUOcelot~\cite{gpuocelot}.
However, in situations where it is not possible to determine the range of memory accessed by the kernel (for example, indirect memory accesses), our approach assumes that the entire GPU memory may be accessed by the kernel.

\vspace{-0.3cm}

\begin{figure}[!htb]
\vspace{-0.1cm}
\begin{lstlisting}  
struct ACE_wrapper {
   //list of read,write segments defined as 
   //[{start_adr1,size1},{start_adr2,size2}..]
   list __read_segments__;
   list __write_segments__;
   // function which gets called at kernel 
   // launch to populate read,write segments
   void get_addresses(
        dim3 blocks, dim3 threads, ...
   );
   // function declaration of the kernel
   static __global__ void kernel(...);
};
\end{lstlisting}
% %\vspace{-0.2cm}
\caption{The \texttt{\Xlabel{}\_wrapper} definition}
\label{fig:kernel_definition}
\vspace{-0.6cm}
\end{figure}

\begin{figure}[!htb]
\vspace{-0.2cm}

\begin{lstlisting}
// get address function for matrix multiply
// input matrices: input1 (mxn), input2(nxk)
// output matrix: output(mxk)
void ACE_wrapper::get_addresses(
    dim3 blocks, dim3 threads,
    int* input1, int* input2, int* output1, 
    int m, int n, int k) {
   // input1 reads m*n elements 
   // input2 reads n*k elements 
   __read_segments__ = {
       {(void*)input1, m*n*sizeof(int)}, 
       {(void*)input2, n*k*sizeof(int)} 
   };
   // output reads m*k elements 
   __write_segments__ = {
       {(void*)output, m*k*sizeof(int)}, 
   };
}
\end{lstlisting}
\vspace{-0.2cm}
\caption{Example: \texttt{get\_addresses} function}
\label{fig:getaddress}
\vspace{-0.2cm}
\end{figure}

\vspace{-0.2cm}
\subsection{\Xlabel{}-SW Design}
% %\vspace{-0.1cm}
\label{sec:swdesign}
\Xlabel{}-SW is implemented as a user-level runtime that is called by the application. The functionalities of \Xlabel{}-SW are performed by multiple independent threads that are launched simultaneously. The \Xlabel{}-SW runtime performs two major tasks: \emph{(i)} implementing and maintaining the scheduling window (window module); and \emph{(ii)} scheduling kernels ready for execution (scheduling module). 

\subsubsection{The window module}
The window module is implemented as a separate thread that manages the input FIFO queue and the scheduling window. All the functionalities of the scheduling window, dependency tracking, and state management are performed in software within this module. This module is called in two ways: First, when a kernel is invoked by the application thread, this module is called and the kernel is inserted into the input queue. Second, the scheduler module (implemented as a separate thread(s)) calls the window module when a kernel completes execution. At this point, the state of \upstream{} lists is updated and the kernel is removed from the scheduling window. The window module constantly  polls the input queue and the scheduling window. When there is a vacancy in the scheduling window and a pending kernel in the input queue, the kernel is moved into the scheduling window. At this point, the window module performs the necessary dependency checks and bookkeeping.
Algorithm~\ref{alg:dependency_checks} describes how the dependency check is performed\deletion{ from the read/write segments}.

\begin{algorithm}[!htb]
\footnotesize
\caption{Dependency check algorithm}
\label{alg:dependency_checks}
\textbf{Input:} $rslist_1$, $wslist_1$, $wslist_2$  \algorithmiccomment{RW segments of scheduling window kernel, w-segment of kernel in inputFIFO} \\ 
\textbf{Output:} $is\_dependent$ \algorithmiccomment{}
\begin{algorithmic}[1]
% \Function{Dependency\_Check($K$, $k\_in$)}{}

\State $is\_dependent = false$ \algorithmiccomment{initial state of  is\_dependent}

\State $rwslist_1 \gets$ $wslist_1 \bigcup rslist_1$    \algorithmiccomment{Read+Write segments}

\ForEach{$segment_1$ in $rwslist_1$}
 \algorithmiccomment{Test for every pair of segments}% \State \Call{clear}{$K$.}
\ForEach{$ws_2$ in $wslist_2$}

 \algorithmiccomment{get start and end virtual memory addresses}
\State $start_1 \gets segment_1.start$ 
\State $end_1 \gets segment_1.start + segment_1.size$
\State $start_2 \gets ws_2.start$
\State $end_2 \gets ws_2.start+ ws_2.size$ 

% \algorithmiccomment{(StartA <= EndB) and (EndA >= StartB)}
 \algorithmiccomment{check overlaps between start and end addresses}

\If{$start_1 < end_2$ and $ end_1 > start_2$}
    \State $is\_dependent = true$ \algorithmiccomment{}
\EndIf

\EndForEach
\EndForEach

% \EndFunction{}
\end{algorithmic}
\end{algorithm}

\subsubsection{The scheduler module}
\begin{figure}[!htb]
\vspace{-0.3cm}
    \centering
    \includegraphics[width=.57\linewidth]{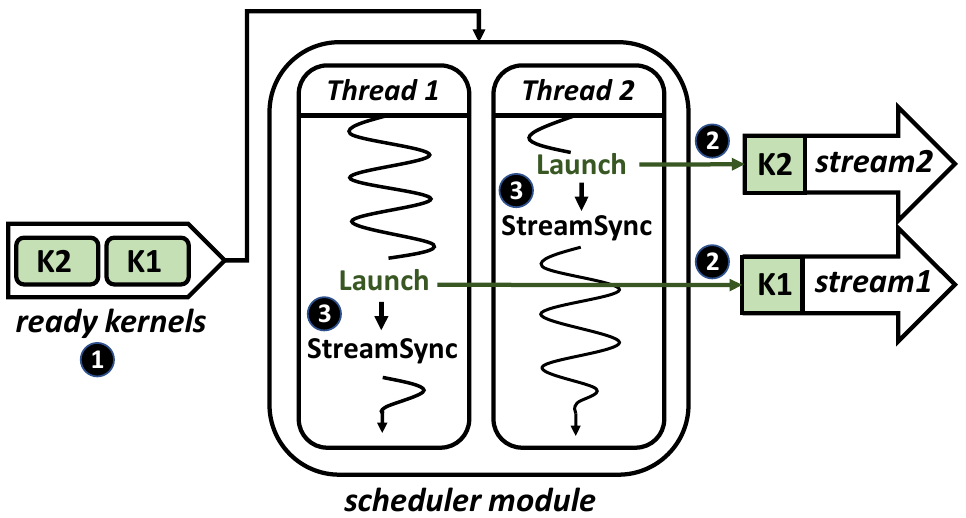}
    % %\vspace{-0.1cm}
    \caption{\Xlabel{}-SW: The scheduler module}
    \label{fig:sw_scheduler}
    \vspace{-0.3cm}
\end{figure}

This module schedules and launches \ready{} kernels for execution. This module is implemented as a configurable fixed number of threads, each of which launches kernels into an independent CUDA stream for concurrent execution, as depicted in Fig.~\ref{fig:sw_scheduler}. Each stream contains only one kernel at any given time.
Threads with empty streams poll the scheduling window for a \ready{} kernel~\circled{1}, which is then launched in its CUDA stream~\circled{2}. The thread then waits for the kernel to complete execution using the StreamSync primitive~\circled{3}. Once the kernel completes execution, the thread calls the window module as described above. This algorithm is described in Algorithm~\ref{alg:worksteal}. 

% Our scheduler initializes a thread corresponding to each initialized stream. Each thread in our scheduler polls looking for a ready-to-launch kernel in the scheduling window, and schedules it for run in the corresponding stream. 

    % %\vspace{-0.1cm}

\begin{algorithm}[!htb]
\footnotesize
\caption{The scheduler module in software}
\label{alg:worksteal}
\hspace*{\algorithmicindent} \textbf{Input:} SchedulingWindow SW, stream\_id
% \hspace*{\algorithmicindent} \textbf{Output:}  

\begin{algorithmic}[1]
\While{ \Call{$not stop()$}{} }     \algorithmiccomment{poll for kernels until stop signal}
\State \Call{acquire\_lock}{$SW$} 
\If{ \Call{$SW.ready.exists{}$} }        \algorithmiccomment{check ready kernels}
    \State $kernel \gets $ \Call{$SW.ready.pop()$}{}     \algorithmiccomment{get ready kernel}
\EndIf
\State \Call{release\_lock}{$SW$}

\State \Call{launch}{kernel, stream\_id} \algorithmiccomment{launch kernel}
\State \Call{stream\_sync}{stream\_id} \algorithmiccomment{wait for completion}
% \State \Call{on\_completion($SW$, $kernel$)}{}
\EndWhile
\end{algorithmic}
\end{algorithm}

%\vspace{-0.2cm}
\subsection{\Xlabel{}-HW Design}
% %\vspace{-0.2cm}
\label{sec:hwdesign}
While \Xlabel{}-SW enables concurrent execution of kernels and can be fully realized in software, it still incurs overheads from \emph{(i)} synchronization with the CPU when a kernel completes execution, i.e., the StreamSync primitive that blocks the scheduler module thread; and \emph{(ii)} the kernel launch overhead when the scheduler module launches a kernel in the CPU. \Xlabel{}-HW is designed to alleviate these overheads with hardware support for kernel scheduling in the GPU\deletion{ that reduces synchronization with the CPU}.

\begin{figure}[!htb]
    \vspace{-0.1cm}
    \centering
    \includegraphics[width=.62\linewidth]{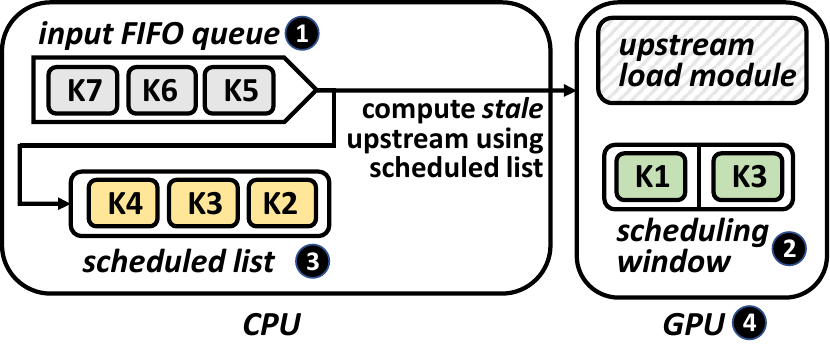}
    % %\vspace{-0.1cm}
    \caption{\Xlabel{}-HW: Design overview}
    \label{fig:hwswdesign}
    \vspace{-0.3cm}
    % %\vspace{-5pt}
\end{figure}

Fig.~\ref{fig:hwswdesign} depicts an overview of \Xlabel{}-HW. \Xlabel{}-HW comprises a software runtime system similar to \Xlabel{}-SW that maintains an input FIFO queue containing the kernels that were invoked by the application \circled{1}. The scheduling window and its management are however implemented in hardware on the GPU side  \circled{2}. The input queue is essentially implemented as a CUDA stream that dispatches kernels to the GPU. In addition to the input FIFO queue, the software runtime also maintains a list of kernels in the GPU's scheduling window, which we call the \scheduledlist{} \circled{3}. To avoid frequent synchronization between the CPU and GPU, we allow this list to be stale. Before a kernel is inserted into the scheduling window, the software runtime performs dependency checks with the 
\scheduledlist{} to determine the upstream kernels. Note that since the \scheduledlist{} may be stale, this upstream list needs to be further updated before insertion into the scheduling window (discussed below).  

    %\vspace{-0.3cm}

The hardware component~\circled{4} consists of two modules: \emph{(i)} the \schedulingwindowhw{} and \emph{(ii)} the \upstream{} load module.

\begin{figure}[!htb]
    % \vspace{-0.3cm}
    \centering
    \includegraphics[width=.75\linewidth]{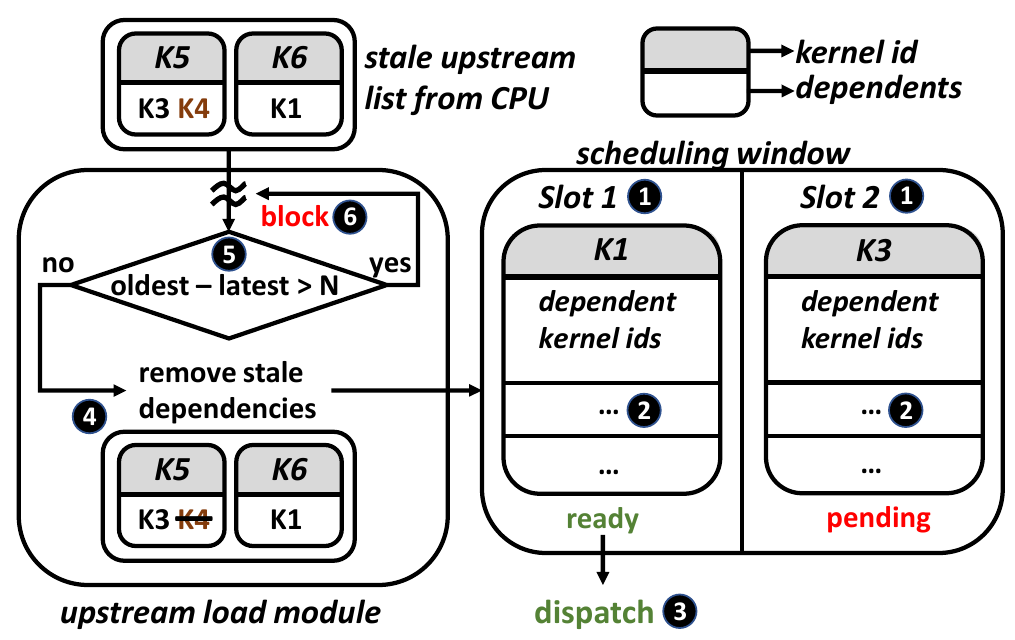}
    % \includesvg[width=.97\linewidth]{graphs_sheets/P3f16.svg}
    % \vspace{-0.1cm}
    \caption{HW scheduling window and upstream load module}
    \label{fig:hwschedulingwindow}
    % \vspace{-0.2cm}
\end{figure}

\textbf{The hardware scheduling window} structure is depicted in Fig.~\ref{fig:hwschedulingwindow} and comprises a fixed number of slots (N)~\circled{1}. Each slot contains an 8-bit kernel identifier and (N-1) 8-bit upstream kernel identifiers that are implemented with SRAM~\circled{2}. \hpcarebuttal{Each slot of the SRAM module is implemented as a single bank of SRAM, contaning N-1 fully associated units to store upstream kernel identifiers.} These upstream identifiers are used to determine when a kernel is \ready{}. 
An additional two bits are used to identify the state of each kernel (i.e., \ready{}, \pending{}, and \executing{}).
When a kernel completes execution, the upstream identifiers are updated and the corresponding state of each kernel is updated. 
The completed kernel is also removed from the scheduling window. 
Any kernels that are now \ready{} are then dispatched to the GPU's kernel dispatch unit for execution~\circled{3}.  

% \vspace{-0.2cm}
    % \vspace{-0.2cm}

\textbf{The upstream load module} is responsible for refining the upstream list provided by the CPU which may be stale in two ways. It may contain kernels that have (1) already completed execution and (2) may miss long-running kernels that are still executing. The first case is handled by the upstream module by checking against a list of kernels in the scheduling window~\circled{4}. The second case is avoided by ensuring that the \scheduledlist{} (of size M) in the CPU never misses kernels that are still executing. The upstream load module tracks the oldest scheduled kernel~\circled{5}. If the number of newer kernels exceeds M (size of the \scheduledlist{}), this module blocks the insertion of more kernels from the CPU~\circled{6}. 

%\vspace{-0.3cm}
\subsection{\Xlabel{} Overheads}
\label{sec:depcheck}
\textbf{(1) Hardware area overhead.}
\Xlabel{}-HW introduces the hardware scheduling window which contains $N$ slots, where $N$ is the size of the scheduling window. Each slot contains $N$ kernel ids of \upstream{} data of 8 bytes each and 2 bits for status. Assuming a scheduling window of length $N=32$, we require 1KB of SRAM for the scheduling module (for the entire GPU).
The upstream module keeps track of the oldest executing kernel with an 8-bit \iscadeletion{\hpcarebuttal{unit of memory in SRAM, and the number of kernels scheduled or completed execution since this oldest executing kernel (\circled{5} in Fig.~\ref{fig:hwschedulingwindow}).} }

\textbf{(2) Storage overheads.} 
The read and write segments that are saved as metadata in the input FIFO and the \scheduledlist{} by the software runtime in the CPU require memory storage. Each read and write segment requires 48 bits to hold the start addresses and the size.

\textbf{(3) Mechanism latencies.} \Xlabel{}-HW requires updating all \upstream{} kernels in each slot of the scheduling window every time a kernel completes execution. \Xlabel{}-HW updates each slot in N-1 cycles (where N is the size of the scheduling window).
Additionally, \Xlabel{}-HW requires $N$ cycles to insert a kernel ID with its \upstream{} kernel IDs into the scheduling window. For a scheduling window of size 64, this operation adds 64 cycles (about $50$-$100ns$) overhead to dispatch a \ready{} kernel for launch. Thus, \Xlabel{}-HW adds negligible runtime to the application compared to the baseline kernel launch overhead (in the order of a few microseconds). 

% Cost of doing dependency checks.
\insertion{\textbf{(4) Dependency checking overheads} 
To determine the list of \upstream{} kernels, the CPU checks for overlaps between the write segments of the kernel in the input queue and the read-write segments of the kernels in the \scheduledlist{}. As the \scheduledlist{} can fit completely into the cache (4KB), dependency-checking is compute-bound and dependent on the number of read and write segments. Table~\ref{tab:overhead} presents the time required to do dependence checking. For a processor with $P$ execution units, effective utilization requires dependency checks to be performed in no more than $T/P$, where $T$ is the task execution time~\cite{task_superscalar,castillo_task_management}. We estimate $T/P$ to be around $4us$, which is much more than the dependency check latency.}

% %\vspace{-0.1cm}
\begin{table}[!htb]
    % %\vspace{-5pt}
    % %\vspace{-0.2cm}
    \centering
    \footnotesize
    \begin{tabular}{|c|c|c|}
\hline
        \textbf{Window} 
        & \textbf{Number of} 
        & \textbf{Dependency}\\

         \textbf{size} 
        & \textbf{RW-segments} 
        & \textbf{check time}\\
\hline
        \multirow{2}{*}{16}  & 6 & 410ns \\
                             & 10 & 700ns \\
\hline
        \multirow{2}{*}{32}  % & \multirow{2}{*}{4us} 
                            & 6 & 510ns \\
                             % &  
                            & 10 & 1640ns \\
\hline
    \end{tabular}
    % \vspace{-0.1cm}
    \caption{\insertion{Dependency checking overhead analysis}}
    \label{tab:overhead}
    \vspace{-0.1cm}
\end{table}

\section{Methodology}
\label{sec:methodology}
\begin{outlineenv}
\outlinepoint{GPU configuration}
We evaluate \Xlabel{}-SW on a real hardware setup with an Intel Core i7 11700K CPU (Table~\ref{table: cpuconfig}) and an NVIDIA RTX3060 GPU (Table~\ref{table: gpuconf}). We model \Xlabel{}-HW on GPUs using the Accel-Sim simulator~\cite{accelsim}, configured with parameters of RTX3070 (Table~\ref{table: gpusim}). We use AccelWattch~\cite{accelsim} to model GPU power. We choose a scheduling window size of $32$.

\begin{table}[!ht]
  \vspace{-0.2cm}
  \centering
  \footnotesize
  \begin{tabular}{|l|}
    \hline
    \hline
     \textbf{CPU} 3.6GHz, OOO 4-wide dispatch window, 32 entry LSQ\\%\\ 128-entry ROB; 32 entry LSQ \\
    \hline
     \textbf{L1D + L1I Cache} 32KB, 4 way LRU, 1 cycle; 64 Byte line; \\ %MSHR size: 10; stride prefetcher \\
    \hline
     \textbf{L2 Cache}  256KB, 8 way LRU, 4 cycle; 64 Byte line; \\ %MSHR size: 10; stride prefetcher \\
    \hline
     \textbf{L3 Cache} 1MB, 16 way LRU, 20 cycle; 64 Byte line; \\ %MSHR size: 64; stride prefetcher \\
    \hline
     \textbf{DRAM} 2-channel; 16-bank; open-row policy, 4GB DDR4\\
    \hline
  \end{tabular}
  \caption{CPU system configuration}
   \vspace{-0.3cm}
  \label{table: cpuconfig}

\end{table}

\begin{table}[!ht]
    %\vspace{-0.5cm}
 \centering
 \footnotesize
  \begin{tabular}{|l|}
    \hline
    \hline
     \textbf{Shader core} 28 SMs, 1.3GHz; 2 schedulers per SM  \\
    \hline
     \textbf{SM Resources} 32768 Registers, 32KB Shared memory, 128KB L1D \\
    \hline
     \textbf{DRAM} 2-channel; 16-bank; open-row policy, 12GB DDR4 \\
    \hline
  \end{tabular}
  \caption{GPU system configuration}
    \vspace{-0.3cm}
  \label{table: gpuconf}
\end{table}

% %\vspace{-9pt}

\begin{table}[!htb]
    \footnotesize
\vspace{-0.4cm}
\vspace{0.4cm}
 \centering
  \begin{tabular}{|l|}
    \hline
    \hline
     \textbf{Shader core} 46 SMs, 1.4GHz; 4 schedulers per SM \\
    \hline
     \textbf{SM Resources} 32768 Registers, 32KB Shared memory, 128KB L1D \\
    % \hline
     % \textbf{Memory Model} FCFS scheduling, 16 banks/channel\\
    \hline
     \textbf{DRAM} 2-channel; 16-bank; open-row policy, 16GB DDR4 \\
    \hline
  \end{tabular}
\caption{Simulated GPU configuration}
\vspace{-0.1cm}
  \label{table: gpusim}
\end{table}

\textbf{Workloads.} We evaluate \Xlabel{} using:

\textbf{(1) Deep RL physics simulations.} Brax~\cite{brax} is a GPU accelerated simulation engine for control tasks in reinforcement learning. We evaluate \Xlabel{} with the Ant (\texttt{ant}), Grasp (\texttt{grasp}), Humanoid (\texttt{human}), Cheetah (\texttt{ct}), and Walker2d (\texttt{w2d}) simulation environments. These environments are MuJoCo~\cite{mujoco} simulations for training RL agents to perform a specific task. For example, \texttt{ant} contains a 3d robot (the agent) with one torso and 4 legs, each with a knee joint, and the goal is to move in a particular direction by controlling its legs. 

\textbf{(2) Dynamic DNNs.} We evaluate our approach for 3 dynamic DNN workloads:
InstaNAS\cite{instanas} (\texttt{I-NAS}) is a dynamic CNN for image classification. We evaluate our approach using the InstaNAS-A architecture on the CIFAR10 dataset. 
% The architecture is instance aware, which means the execution proceeds through a different sequence of convolutions dependent upon the input image.
Dynamic routing~\cite{dynamic_routing} (\texttt{DR}) is a DNN trained for semantic segmentation of images. We evaluate our approach on the Dynamic-A 16 layer architecture using the Cityscapes dataset~\cite{cityscape}.
Conditional Convolution~\cite{condconv} (\texttt{CC}) is a mixture-of-experts CNN model for image classification where the weights of the convolutions are computed at runtime. We evaluate the version of Conditional Convolution with 4 experts that uses an efficientnet b4\cite{efficientnet} network as the backbone. 
All three dynamic DNNs are designed for a batch size of 1 and the input image defines the DNN architecture. We use Pytorch~\cite{pytorch} implementations. 

\textbf{(3) Static DNNs.} CNN architectures optimized for low inference latency using neural architecture search (NAS): NASNet~\cite{nasnet} (\texttt{NASNet}), AmoebaNet~\cite{amoebanet} (\texttt{Amoeba}), SqueezeNet~\cite{squeezenet} (\texttt{Squeeze}), and RandomWire~\cite{randomwire} (\texttt{RW}). These CNNs have highly irregular structures with many small kernels. We evaluate \Xlabel{} with a batch size of 1 on CIFAR10.

%\vspace{-.3cm}

\end{outlineenv}

    % %\vspace{-0.1cm}

\section{Evaluation}
% %\vspace{-0.2cm}
\label{sec:eval}
We evaluate \Xlabel{} using three designs: \insertion{\emph{(i)} Baseline: cuDNN implementation (for DNNs) and a jax implementation~\cite{brax} (for deep RL simulation), both using CUDA streams.} \emph{(ii)} \XlabelSW{}: Our software-only mechanism is evaluated on real hardware. \emph{(iii)}  \XlabelSWSim{}: Our software-only mechanism evaluated on the GPU simulator. We also include these results to compare against \XlabelHW{}. \emph{(iv)} \XlabelHW{}: Our hardware-software cooperative mechanism evaluated on the GPU simulator. \insertion{\emph{(v)} \cudagraph{}: Framework where the inter-kernel dependencies are prepared on the CPU as a directed acyclic graph and sent to the GPU ahead of time.} We only present \XlabelSW{} results for the deep RL workloads as the dynamic and static DNNs heavily use CuDNN libraries that do not currently allow modifications to make use of different CUDA streams. We instead model the same effect with \XlabelSWSim{}.

\subsection{Deep RL Physics Simulations}
% %\vspace{-0.1cm}

\begin{figure}[!htb]
\vspace{-0.2cm}
    \centering
    \includegraphics[width=.75\linewidth]{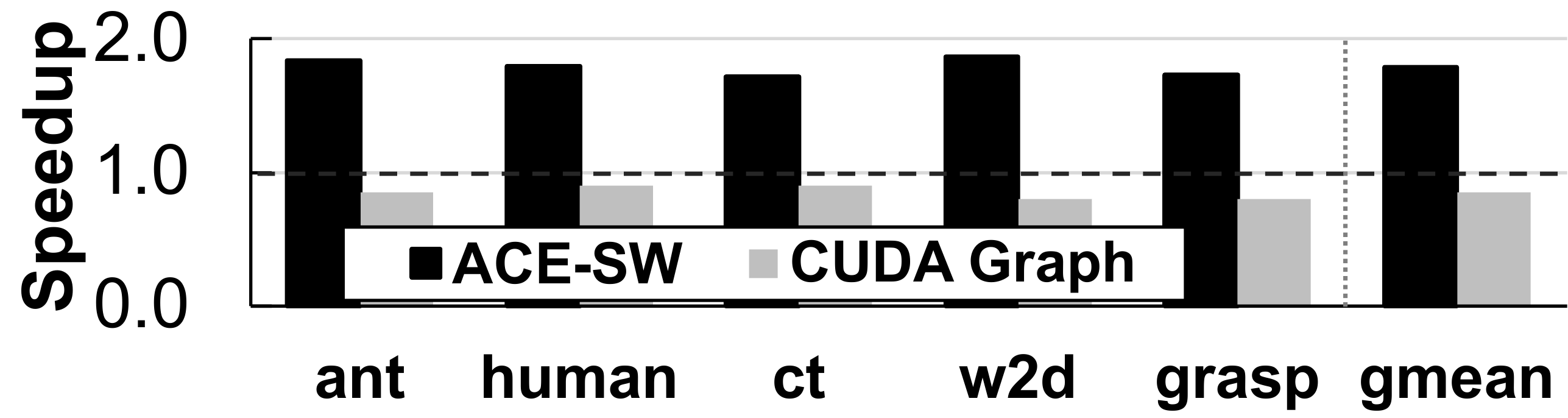}
    % %\vspace{-0.1cm}
    \caption{Deep RL physics simulations: Normalized Speedup}
    \label{fig:brax_speedup}
\vspace{-0.2cm}
\end{figure}

Fig.~\ref{fig:brax_speedup} depicts the runtimes for the generation of a single batch of training data from different simulation environments using \XlabelSW{}, normalized to the baseline approach. 

\iscadeletion{

Table~\ref{tab:speedup_brax_sw} contains the corresponding runtimes. 
We observe that \XlabelSW{} is able to provide performance speedups of up to $1.87\times$ and on average $1.79\times$ over the baseline on real hardware. \XlabelSW{} is thus able to effectively overlap the execution of independent kernels to improve GPU utilization and thus performance.

\begin{table}[!htb]
    \vspace{-0.4cm}
    \centering
    \footnotesize
    \begin{tabular}{|c|c|c|}
\hline
        \textbf{Workload} & \textbf{\XlabelSW{}} & \textbf{baseline} \\
\hline
        \texttt{ant} & 24s & 44s \\
\hline
        \texttt{human} & 40s & 72s \\
\hline
        \texttt{ct} & 18s & 31s \\
\hline
        \texttt{w2d}  & 20s & 38s \\
\hline
        \texttt{grasp}  & 38s & 66s \\
\hline
    \end{tabular}
    % %\vspace{-0.1cm}
    \caption{Deep RL physics simulations: Execution latencies}
    \label{tab:speedup_brax_sw}
\vspace{-0.4cm}
\end{table}
\vspace{-0.2cm}
}

\begin{figure}[!htb]
    % %\vspace{-0.3cm}
    \centering
    \includegraphics[width=.8\linewidth]{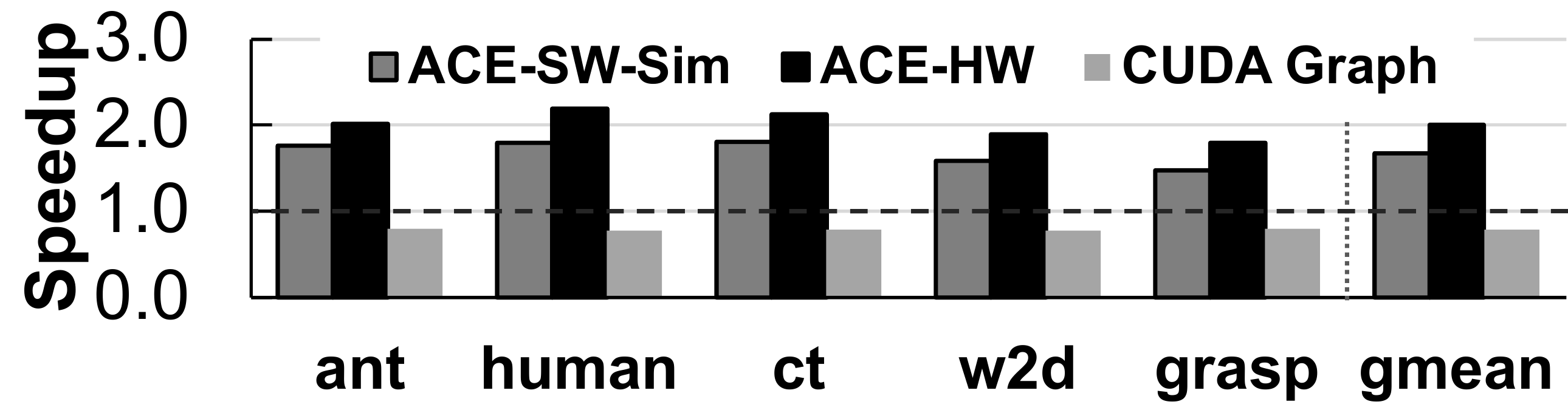}
    % %\vspace{-0.1cm}
    \caption{Deep RL physics simulations: Normalized speedup}
    \label{fig:brax_speedup_sim}
    % %\vspace{-0.3cm}
\end{figure}

Fig.~\ref{fig:brax_speedup_sim} depicts the runtimes for \XlabelSWSim{} and \XlabelHW{} normalized to the baseline implementation. We make two observations. First, \XlabelSWSim{} provides similar speedups as in real hardware compared to the baseline implementation (up to $1.79\times$ and $1.66\times$ on average). Second, \XlabelHW{} is able to further improve performance compared to the software-only approach by alleviating the synchronization and kernel launch overheads. \insertion{We observe a slowdown with \cudagraph{} due to the significant latency of constructing the kernel dependency graph and sending the information to the GPU.}

\begin{figure}[!htb]
    % \vspace{-0.3cm}
    \centering
    \includegraphics[width=.77\linewidth]{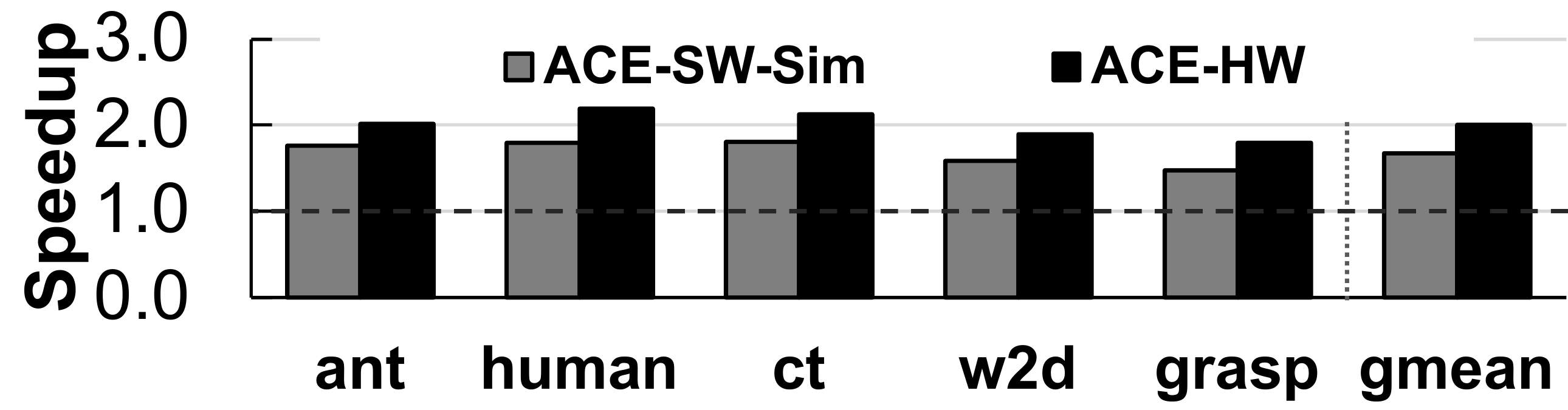}
    % \vspace{-0.1cm}
    \caption{\insertion{Brax: End-to-end speedup}}
    \label{fig:brax_speedup_lrn_sim}
    \vspace{-0.1cm}
\end{figure}

\insertion{The end-to-end speedup in training tasks (simulation + learning algorithm) as  observed is shown in Fig.~\ref{fig:brax_speedup_lrn_sim}. We observe a mean speedup of $1.42\times$ on \XlabelHW{}, and $1.30\times$ on \XlabelSW{}.}

    % \vspace{-0.1cm}

\begin{figure}[!htb]
    % \vspace{-0.2cm}
    \centering
    \includegraphics[width=.8\linewidth]{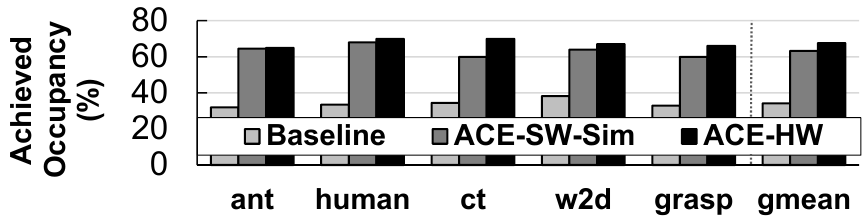}
    % \includegraphics[width=1.\linewidth]{graphs_new_svg/brax_occupancy2.pdf}
    % \vspace{-0.1cm}
    \caption{Deep RL physics simulations: Achieved occupancy}
    \label{fig:brax_occupancy}
    \vspace{-0.3cm}
\end{figure}

In Fig.~\ref{fig:brax_occupancy}, we depict the achieved occupancy for the three configurations. Achieved occupancy is calculated as the number of active warps divided by the maximum number of active warps supported by the GPU averaged over all clock cycles. We observe that the \Xlabel{} is able to significantly increase the achieved occupancy and thus the utilization.

%\vspace{-0.2cm}
\subsection{Inference on Dynamic DNNs}
\begin{figure}[!htb]
    \vspace{-0.2cm}
    \centering
    \includegraphics[width=0.65\linewidth]{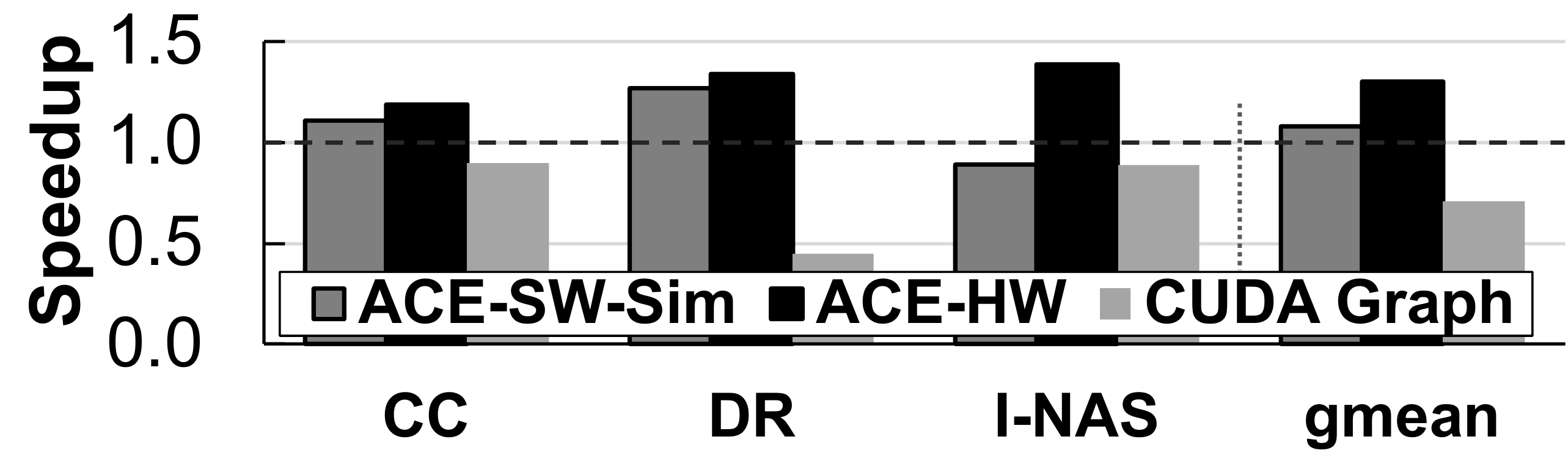}
    \vspace{-0.1cm}
    \caption{\insertion{Dynamic DNNs: Normalized speedup}}
    \label{fig:speedup_dynconv}
    \vspace{-0.3cm}
\end{figure}

Fig.~\ref{fig:speedup_dynconv} depicts speedup over the baseline for the dynamic DNNs described in~\cref{sec:methodology}. We observe that \Xlabel{} is able to provide speedups of up to $1.39\times$ on dynamic DNN workloads with \XlabelHW{} and on average $1.05\times$ with \XlabelSW{} and $1.3\times$ with \XlabelHW{}. 
\texttt{I-NAS} suffers a slowdown with \XlabelSW{} because this workload has significant kernel launch overheads when parallelized but are hidden in the baseline case where the kernels are simply launched serially into a single stream without synchronization. \insertion{We observe that \cudagraph{} exhibits a significant slowdown due to the overhead incurred during the construction and communication of the DAG dependencies. }

\begin{figure}[!htb]
    \vspace{-0.3cm}
    \centering
    \includegraphics[width=0.7\linewidth]{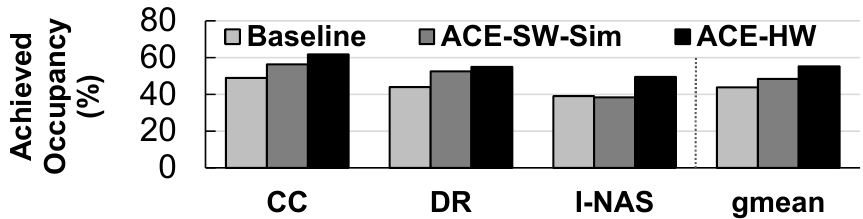}
    \caption{Dynamic DNNs: Achieved occupancy}
    \vspace{-0.2cm}
    \label{fig:dynconv_occupancy}
    % \vspace{-5pt}
    \vspace{-0.4cm}
\end{figure}

Fig.~\ref{fig:dynconv_occupancy} depicts the corresponding achieved occupancy. We find that the \Xlabel{} configurations are able to significantly improve utilization, leading to performance improvements.  

\vspace{-0.1cm}

\subsection{Inference on Static DNNs}
\begin{figure}[!htb]
    % \vspace{-0.3cm}
    \centering
    \includegraphics[width=.75\linewidth]{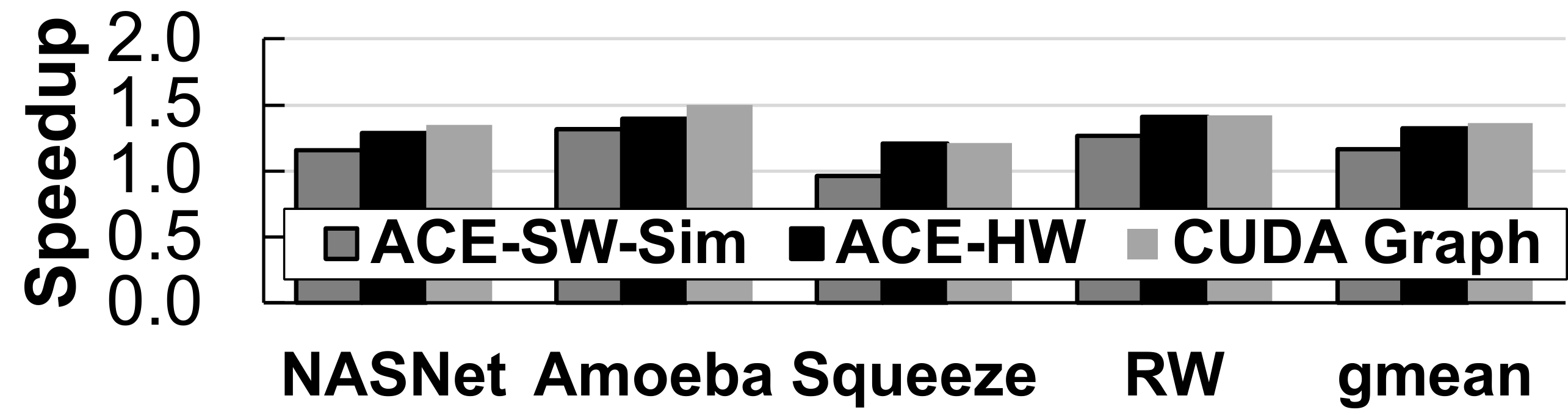}
    % \vspace{-0.1cm}
    \caption{\insertion{Static DNNs: Normalized speedup}}
    \label{fig:speedup_nas}
    % \vspace{-0.3cm}
\end{figure}

\begin{figure}[!htb]
    % \vspace{-0.3cm}
    \centering
    \includegraphics[width=.7\linewidth]{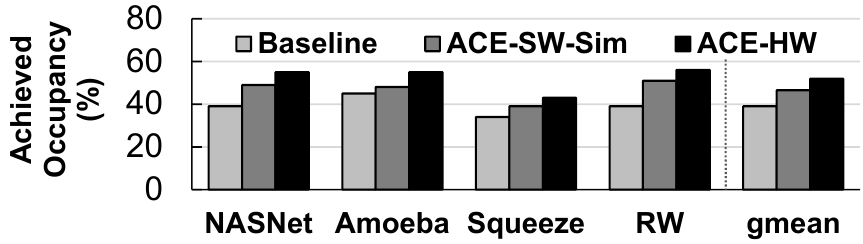}
    % \vspace{-0.1cm}
    \caption{Static DNNs: Achieved occupancy}
    \label{fig:occupancy_nas}
    % \vspace{-0.3cm}
\end{figure}

%\vspace{-0.1cm}
While our approach is designed for applications with dynamic computational graphs, we also evaluate its effectiveness in improving the concurrency of static DNNs. We depict the speedups obtained normalized to the baseline in Fig.~\ref{fig:speedup_nas}. We observe an average speedup of $1.31\times$ with \XlabelHW{}, and a speedup of $1.16\times$ with \XlabelSW{}. 
Fig.~\ref{fig:occupancy_nas} depicts the corresponding achieved occupancy. We find that \Xlabel{} leads to higher GPU utilization, leading to performance improvements. \insertion{As expected, we observe that \cudagraph{} exhibits similar execution times as \XlabelHW{} for static graphs. This is because the task graph needs to be constructed only once.
}

\subsection{Sensitivity Analysis}
\begin{figure}[!htb]
    \vspace{-0.3cm}
    \centering
    \includegraphics[width=.75\linewidth]{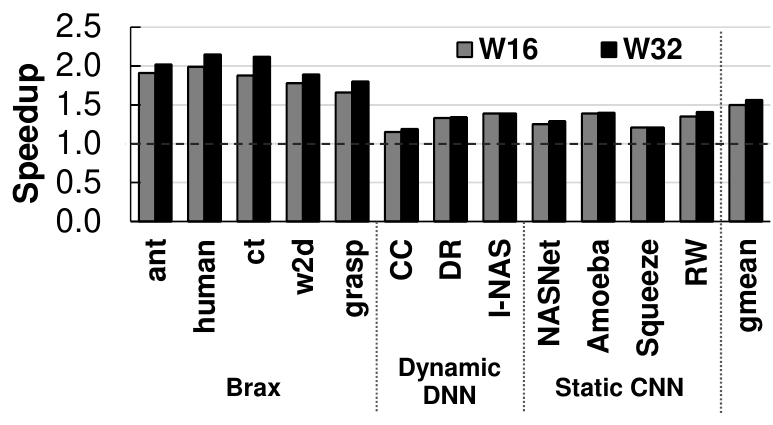}
    \vspace{-0.2cm}
    \caption{\insertionhpca{Speedups on varying scheduling window size}}
    \label{fig:sched_window_change}
    \vspace{-0.2cm}
\end{figure}

% %\vspace{-0.2cm}
Fig.~\ref{fig:sched_window_change} compares the speedups obtained on using scheduling window sizes of 16 and 32 for \XlabelHW{} over baseline. We observe that the Brax simulations have higher performance (4.5\% on average) with a window size of 32 compared to 16. However, the window size has less of an impact on the DNNs. This is because the simulation engines have more inter-kernel parallelism that is exposed with a larger scheduling window. 

\vspace{-0.2cm}
    \vspace{-0.3cm}

\iscadeletion{
\subsection{Energy Consumption}
\vspace{-0.2cm}
Fig.~\ref{fig:energy_consumption_reduction} shows the normalized energy consumption for the evaluated workloads. We observe energy savings of 21.6\% on average on all workloads with \XlabelHW{}, and 6.1\% with \XlabelSW{} as a result of the reduction in execution time.

\vspace{-0.1cm}
\begin{figure}[!htb]
    \vspace{-0.2cm}
    \centering
    \includegraphics[width=.75\linewidth]{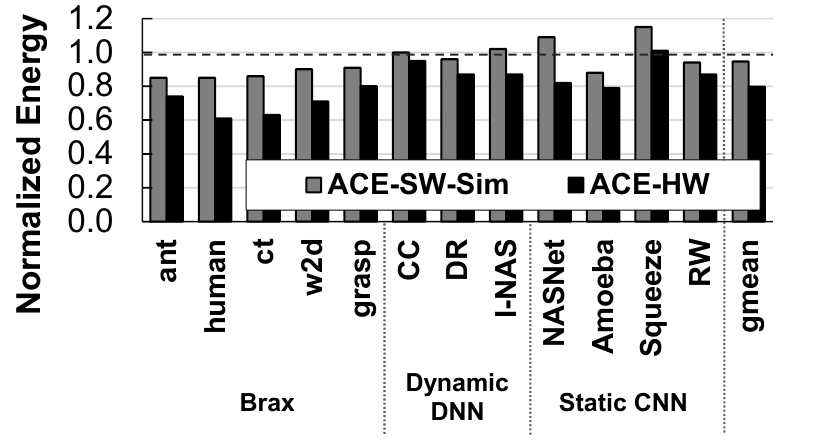}
    \vspace{-0.1cm}
    \caption{Normalized energy consumption}
    \label{fig:energy_consumption_reduction}
    \vspace{-0.3cm}
\end{figure}
    \vspace{-0.2cm}
}
\subsection{Comparison with Persistent Thread Frameworks}
\vspace{-0.1cm}
\label{sec:persistent_threads}
\insertionasplos{Persistent threads (PT)~\cite{juggler, atos, pt_study, pt_raytraversal} are used to efficiently schedule multiple tasks with dynamically determined dependencies. These tasks are executed using threads of a \emph{single kernel}. Thus, it assumes all tasks are \emph{homogeneous}, requiring the same number of registers and shared memory. PT frameworks which allow heterogeneous kernels are non-trivial and would be inefficient as the persistent kernel must be configured to use the maximum registers/scratchpad used by any kernel~\cite{atos}.
We use the persistent thread framework implementation from juggler~\cite{juggler} and adapted it to handle heterogeneous kernels. We were only able to implement a section of a rigid body simulator (used for finding contacts between pairs of rigid bodies). %The code iterates through every pair of objects to check for contacts between them using a collision detection routine. 
This routine invokes a different kernel (with different register usages) for different pairs of geometries. We implement these kernels as tasks of our PT framework and find that it is $1.35\times$ slower than baseline. This slowdown is due to inefficient use of registers/scratchpad by the  kernel that leads to lower parallelism.}
% }
%\vspace{-0.2cm}

%low utilization observed in some tasks, as we have set the registers to be used by each thread to the maximum required by all tasks.}

% \deletion{
% \begin{figure}[!htb]
% %\vspace{-0.5cm}
% \begin{lstlisting}
% // Routine to find contacts between 
% // bodies in the system
% // input: system sys
% // output: list of contacts c
% def find_contacts(sys):
%     for geom1, geom2 in sys.bodies:
%         // get the corresponding collision
%         // detection function for types of geometries
%         colldet_fn = get_fun(geom1.type, geom2.type)
%         c = colldet_fn(geom1, geom2)
%     return c
% \end{lstlisting}

% \caption{Benchmark function to compare with PT frameworks}
% \label{fig:pt_example}
% %\vspace{-0.5cm}
% \end{figure}
% }

\section{Related Work}
% \vspace{-0.1cm}
\label{sec:related_work}
% To our knowledge, this is first work to
\insertionhpca{In this work, we}(i) observe that input-dependent inter-kernel dependencies and small kernels are a significant performance bottleneck in a range of important applications such as simulation engines in deep RL and dynamic neural networks; and (ii) propose both a software-only and hardware-software cooperative mechanism to enable concurrent execution of kernels with statically unknown inter-kernel dependencies. In this section, we describe prior work that aim to improve GPU utilization and kernel concurrency. 

\textbf{Leveraging concurrent streams in DL workloads.}
Mainstream deep learning frameworks like Tensorflow~\cite{tensorflow} and Pytorch~\cite{pytorch} launch GPU kernels into a single CUDA stream that executes them sequentially. 
Recent works~\cite{ooobackprop,ios,daydream} propose software techniques to enable concurrent execution of GPU kernels using multiple streams with static scheduling and stream assignment before application execution.
Inter-operator scheduling~\cite{ios} partitions a computation graph into sections of kernels that can execute in parallel.
Out-of-order backprop~\cite{ooobackprop} observes that gradient computation can be parallelized using CUDA streams into weight gradients and the output gradient computation during backpropagation.
However, these works are only applicable to DL workloads whose computation graph is static and known ahead of time, often requiring significant compilation times.
Furthermore, these approaches incur high synchronization overheads.

\insertionasplos{
\textbf{Task-based programming frameworks in CPUs.}
Task-based frameworks~\cite{onetbb, cilk, openmp} enable programmers to describe a program as multiple tasks which are scheduled for execution in multiprocessor architectures~\cite{sarc}. Works such as task superscalar~\cite{task_superscalar}, carbon~\cite{carbon}, TDM~\cite{castillo_task_management} and ADM~\cite{adm} propose out-of-order scheduling of tasks to efficiently leverage irregular parallelism in multiprocessors.
The major bottleneck in out-of-order scheduling of tasks dynamically for multiprocessors is the long latency required to do dependence checks. Thus, prior work~\cite{task_superscalar, castillo_task_management,adm, carbon} propose hardware accelerators to address the long latency dependence checking needed at runtime. 
\iscadeletion{
\hpcarebuttal{Detecting data dependencies during runtime by inspecting memory overlaps between various program data can pose a bottleneck when scheduling computations in multiprocessors. To enable a superscalar, out-of-order dataflow execution model for \emph{statically-sequential} programs on multi-core processors, efficient dependence checks can be done by checking for data dependencies between objects in the program instead of individual memory locations for efficient~\cite{multicoredataflow}. Serialization sets~\cite{serialization_sets} unlock parallelism in the application by determining the data read to/written by each task data at different stages of the program execution, and scheduling independent tasks concurrently on a multiprocesor.}}
However, with GPUs, the primary bottleneck is the long latency to launch/signal completion of kernels instead, requiring a different approach to enable out-of-order scheduling.

\textbf{Programmer annotations} Prior works leverage programmer-specified annotations as hints to the compiler to extract parallelism.
DeNovo~\cite{denovo} uses programmer annotations that encode the data read and written to by each method/function. This information is used at compile time to determine independent tasks that can be scheduled. % similar but applied contexts
Some frameworks~\cite{starss,openstream,openmp, multicoredataflow, serialization_sets} allow programmers to annotate the array regions accessed by each task as a compile time directive.  
%StarSs~\cite{starss}, OpenStream~\cite{openstream}, and OpenMP tasks~\cite{openmp} allow programmers to annotate the array regions accessed by each task as a compile time directive in the program.  
In \Xlabel{}, we use a similar approach of programmer annotations to help determine parallelism at runtime to enable out-of-order kernel scheduling.}

% \insertionasplos{
% \textbf{Use of programmer annotations to extracting further parallelism} Prior works such as StarSs~\cite{starss}, OpenStream~\cite{openstream}, and OpenMP tasks~\cite{openmp} allow programmers to annotate the array regions accessed by each task as a compile time directive in the program. 
% DeNovo~\cite{denovo} uses programmer annotations that encode the data members' (objects in the program) read and written to by each method/function to determine independent tasks that could be scheduled with no data race. In \Xlabel{}, we adopt a similar approach by leveraging programmer-specified annotations as hints for the runtime to extract parallelism between kernels. However, we go a step further by demonstrating how this extracted parallelism can be harnessed to enable faster execution with out-of-order kernel execution.}

\deletion{\textbf{Task-based programming frameworks in CPUs.}
Task-based frameworks~\cite{onetbb, cilk, openmp} enable programmers to describe a program as multiple "tasks" which are scheduled for execution. Prior works leverage similar programmer-specified annotations as hints to the compiler to extract further parallelism. For example, the DeNovo project~\cite{denovo} uses programmer annotations that encode the data members' (of objects in the program) read and written to by each method/function. This information is used at compile time to determine independent tasks that could be scheduled with no data race. % similar but applied contexts
StarSs~\cite{starss}, OpenStream~\cite{openstream}, and OpenMP tasks~\cite{openmp} allow programmers to annotate the array regions accessed by each task as a compile time directive in the program. 
One major bottleneck in out-of-order scheduling of tasks dynamically for CPU multiprocessors is the long latency required to do dependence checks. Task superscalar~\cite{task_superscalar}, TDM~\cite{castillo_task_management}, ADM~\cite{adm} and carbon~\cite{carbon} propose hardware accelerators to address the long latency dependence checking. However, with GPUs, the primary bottleneck is the long latency required to launch and signal completion of kernels instead, requiring a different approach to enable out-of-order kernel scheduling.}

\textbf{Software techniques to improve GPU utilization with concurrent kernel execution.}
CUDA Graphs~\cite{cudagraph} and AMD ATMI~\cite{atmi, dagee, hipgraph} are frameworks that allow users to define dependencies between kernels as a directed-acyclic-graph (DAG) prior to execution. \deletion{ATMI~\cite{atmi} utilizes HSA barrier packets~\cite{hsa_barrier, gpu_command_queues}, to express data dependencies between kernels. It blocks the launch of kernels until their data dependencies are met.}
This approach eliminates synchronization and kernel launch overheads due to communication with the CPU. 
Nimble~\cite{nimble} identifies independent GPU kernels prior to execution and concurrently schedules independent kernels using CUDA streams. This approach uses CUDA Graphs~\cite{cudagraph} to reduce synchronization and kernel launch overheads.
\insertionasplos{Irregular graphs are also seen in solving sparse linear equations for CFD simulations~\cite{ata} and hyperplane sweep routines~\cite{hyperplane_sweep_optimization}, where DAG frameworks have been shown to be effective.}\deletion{ ATA~\cite{ata} leverages ATMI~\cite{atmi} to express the irregular task graphs seen in solving sparse linear equations for CFD simulations. Kaushik et. al,~\cite{hyperplane_sweep_optimization} used ATMI~\cite{atmi} to express the dependencies between operations in hyperplane sweep algorithm to leverage kernel concurrency.}\deletion{AMD's open-source frameworks~\cite{hipgraph} have leveraged ATMI into several graph analytics and learning applications.}\insertion{We quantitatively compared \Xlabel{} against a CUDA graph implementation in~\cref{sec:eval}.}
None of these approaches is applicable to dynamic input-dependent computational graphs, as caching dependency information and constructing CUDA Graphs incur non-trivial latencies (~\cref{sec:motivation_cudagraph}). \iscadeletion{\insertionasplos{Persistent threads~\cite{juggler, whippletree, atos, pt_study, pt_raytraversal} allow efficient scheduling of tasks with dynamic data dependencies between them. We discuss persistent threads in ~\cref{sec:persistent_threads}.}}

\iscadeletion{GPU frameworks such as multi-instance GPU (MIG)~\cite{mig} or multi-processing service (MPS)~\cite{mps} offered by CUDA allow increase in the computation throughput for \emph{multiple} jobs in data centers, through GPU sharing. However, the focus of our work is to address the bottlenecks associated with workloads consisting of several short-running kernels with dynamically determined inter-kernel dependencies for a single application. Cooperative thread groups offer a convenient interface to do synchronizations and data movements across a group of threads at a fine-grained level. This allows for the implementation of intra-kernel optimizations. However, intra-kernel optimization is orthogonal to the issue addressed by \Xlabel{} in which we address the problem  of extracting parallelism across GPU kernels and efficiently scheduling them.
}

\deletion{,making this an infeasible approach to be done at runtime.} \deletion{We also note that CUDA graphs require the movement and storage of large graph data to the GPU.}
\deletion{Persistent thread (PT) frameworks~\cite{juggler, atos, pt_study} are used to efficiently schedule multiple tasks with dynamically determined dependencies. These tasks are executed using threads from a \emph{single kernel} that is launched at the program start. Thus, it assumes all kernels/thread blocks are \emph{homogeneous}, requiring the same number of registers, shared memory, and thread blocks. It is therefore challenging to map deepRL and dynamic NN workloads that comprise \emph{heterogeneous} kernels with irregular dependencies as tasks in a PT framework. Modifying persistent thread frameworks to allow heterogeneous kernels is non-trivial and would be inefficient as the persistent kernel must be configured to use the maximum registers/scratchpad used by any kernel.}

\textbf{Hardware support for concurrent kernels.}
Wireframe~\cite{wireframe} proposes merging multiple kernels into a single large kernel and performs CTA scheduling with data dependency checks between CTAs.
Blockmaestro~\cite{blockmaestro} enables concurrently running kernels by identifying dependencies between their CTAs. 
These approaches however perform dependence checks by tracing and extracting the memory loads and stores performed by each thread block of every kernel. Similar to the software approaches, these approaches are designed for static computational graphs. The proposed scheduling and dependency check techniques would be too time-consuming for runtime scheduling.
%Thus they are not applicable to workloads with runtime-determined computational graphs.
\insertionasplos{GPU dynamic parallelism~\cite{cdp,free_launch, klap, dynamic_tb_launch} enables launching kernels from the device itself and allows data dependencies between a single parent and multiple child kernels. However, Dynamic-NN and RL simulation workloads contain kernels that depend on multiple kernels, making it difficult to apply GPU dynamic parallelism.}

\textbf{Compilers, runtime systems for dynamic neural networks.} % Similar to \Xlabel{}-SW, PTask~\cite{ptask}, Dandelion~\cite{dandelion}
Prior software~\cite{cortex, janus, cavs, nimble_dynamicnn, dynamic_nn_need, deepframeworks_recursion, dl_dynamic_comp_basic} and hardware approaches~\cite{parameter_caching} aim to optimize CPU-GPU communication overheads, launch overheads, and blocking synchronization calls for dynamic computational graphs. 
These approaches introduce techniques such as dynamic batching and kernel fusion. However, these works are orthogonal to our approach. 
\insertionhpca{Prior works~\cite{ptask, dandelion} have proposed software frameworks for CPU-GPU systems that provide simplified and convenient abstractions to interface with GPU runtime APIs. These frameworks encapsulate runtime-level code, simplifying code development for programmers in single and multi-GPU environments. However, these works do not specifically focus on input-dependent dynamic computation. Instead their goal is to provide simpler abstractions for programming GPU tasks and expressing dataflow dependencies between them.
Efficient GPU sharing techniques, such as Kernelet~\cite{kernelet}, GPUPool~\cite{gpupool} introduce runtime systems to enable concurrent kernel execution by scheduling kernels from different processes which have different memory and compute usage intensities. However, while these works increase overall GPU utilization by kernels launched from different processes, they do not leverage the parallelism between kernels of a single application.}

%, as thdo not address GPU resource underutilization due to small kernels. 

\vspace{-0.4cm}

% \section{Implementation Notes and Limitations}
% \input{textsrcs/implementation_notes}

\section{Conclusion}
% \vspace{-0.2cm}
We introduce \Xlabel{}, the first framework that enables automatic concurrent kernel execution with low overhead runtime scheduling and dependency checks. The key idea behind \Xlabel{} is to dynamically schedule a small window of kernels by identifying which kernel(s) within the window is ready for execution. \Xlabel{} leverages kernel annotations to automatically identify kernel dependencies at runtime. We implement \Xlabel{} as both a software framework and a hardware-software mechanism that is able to further reduce synchronization overheads from CPU-GPU communication. We demonstrate that \Xlabel{} can improve the performance of important emerging classes of workloads, such as RL simulations and dynamic DNNs, whose kernel dependencies are irregular and vary with input.

% \pagebreak{}

%In this paper, we present 1) \Xlabel{}-SW, a software technique and 2) \Xlabel{}-HW, a software-hardware mechanism to schedule GPU kernels concurrently while tracking inter-kernel data dependencies at runtime. 
%Our \Xlabel{}-SW mechanism \todo{} %The key idea of our approach.
%With our approach, we demonstrate a speedup of up to $\times$ on GPU accelerated RL training applications and CNN inference, which have irregular data dependencies between them at runtime.
%We develop \Xlabel{}-HW, a lightweight hardware-software mechanism to further reduce the  by eliminating overhead of synchronization and GPU kernel launch overhead. The key idea of \Xlabel{}-HW is to in global memory
%With \Xlabel{}-HW, we demonstrate a speedup of up to $\times$ on irregular data dependencies between them GPU accelerated RL-training and  

% \bibliographystyle{plain}
\bibliographystyle{ieeetr}
\bibliography{references}

\begin{thebibliography}{100}

\bibitem{brax}
C.~D. Freeman, E.~Frey, A.~Raichuk, S.~Girgin, I.~Mordatch, and O.~Bachem, ``Brax - a differentiable physics engine for large scale rigid body simulation,'' {\em ArXiv}, vol.~abs/2106.13281, 2021.

\bibitem{isaacsim}
V.~Makoviychuk, L.~Wawrzyniak, Y.~Guo, M.~Lu, K.~Storey, M.~Macklin, D.~Hoeller, N.~Rudin, A.~Allshire, A.~Handa, and G.~State, ``Isaac gym: High performance gpu-based physics simulation for robot learning,'' {\em ArXiv}, vol.~abs/2108.10470, 2021.

\bibitem{large_batch_drl}
B.~Shacklett, E.~Wijmans, A.~Petrenko, M.~Savva, D.~Batra, V.~Koltun, and K.~Fatahalian, ``Large batch simulation for deep reinforcement learning,'' {\em ArXiv}, vol.~abs/2103.07013, 2021.

\bibitem{atari_gpu}
S.~Dalton and I.~Frosio, ``Accelerating reinforcement learning through gpu atari emulation,'' {\em arXiv: Learning}, 2020.

\bibitem{samplefactory}
A.~Petrenko, Z.~Huang, T.~Kumar, G.~Sukhatme, and V.~Koltun, ``Sample factory: Egocentric 3d control from pixels at 100000 fps with asynchronous reinforcement learning,'' in {\em ICML}, 2020.

\bibitem{ddw}
K.~Yuan, Q.~Li, S.~Guo, D.~Chen, A.~Zhou, F.~Yu, and Z.~Liu, ``Differentiable dynamic wirings for neural networks,'' {\em 2021 IEEE/CVF International Conference on Computer Vision (ICCV)}, pp.~317--326, 2021.

\bibitem{d2nn}
L.~Liu and J.~Deng, ``Dynamic deep neural networks: Optimizing accuracy-efficiency trade-offs by selective execution,'' in {\em AAAI}, 2018.

\bibitem{s2dnas}
Z.~Yuan, B.~Wu, Z.~Liang, S.~Zhao, W.~Bi, and G.~Sun, ``S2dnas: Transforming static cnn model for dynamic inference via neural architecture search,'' {\em ArXiv}, vol.~abs/1911.07033, 2020.

\bibitem{dynamic_nn_survey}
Y.~Han, G.~Huang, S.~Song, L.~Yang, H.~Wang, and Y.~Wang, ``Dynamic neural networks: A survey,'' {\em IEEE Transactions on Pattern Analysis and Machine Intelligence}, vol.~44, pp.~7436--7456, 2022.

\bibitem{instanas}
A.~Cheng, C.~H. Lin, D.-C. Juan, W.~Wei, and M.~Sun, ``Instanas: Instance-aware neural architecture search,'' in {\em AAAI}, 2020.

\bibitem{dynamic_nn_need}
J.~Wei, G.~Gibson, V.~Vasudevan, and E.~Xing, ``Dynamic scheduling for dynamic control flow in deep learning systems,'' {\em URL http://www. cs. cmu. edu/jinlianw/papers/dynamic\_scheduling\_nips18\_sysml. pdf}, 2018.

\bibitem{dynamic_routing}
S.~Cai, Y.~Shu, and W.~Wang, ``Dynamic routing networks,'' {\em 2021 IEEE Winter Conference on Applications of Computer Vision (WACV)}, pp.~3587--3596, 2021.

\bibitem{rdinet}
H.~Wang, S.~Li, S.-C. Su, Z.~Qin, and X.~Li, ``Rdi-net: Relational dynamic inference networks,'' {\em 2021 IEEE/CVF International Conference on Computer Vision (ICCV)}, pp.~4601--4610, 2021.

\bibitem{skipconv}
P.~Singh and V.~P. Namboodiri, ``Skipconv: skip convolution for computationally efficient deep cnns,'' in {\em 2020 International Joint Conference on Neural Networks (IJCNN)}, pp.~1--8, IEEE, 2020.

\bibitem{branchynet}
S.~Teerapittayanon, B.~McDanel, and H.~T. Kung, ``Branchynet: Fast inference via early exiting from deep neural networks,'' {\em 2016 23rd International Conference on Pattern Recognition (ICPR)}, pp.~2464--2469, 2016.

\bibitem{blockdrop}
Z.~Wu, T.~Nagarajan, A.~Kumar, S.~Rennie, L.~S. Davis, K.~Grauman, and R.~Feris, ``Blockdrop: Dynamic inference paths in residual networks,'' in {\em CVPR}, 2018.

\bibitem{convaig}
A.~Veit and S.~J. Belongie, ``Convolutional networks with adaptive inference graphs,'' {\em International Journal of Computer Vision}, vol.~128, pp.~730--741, 2019.

\bibitem{micronet}
Y.~Li, Y.~Chen, X.~Dai, D.~Chen, M.~Liu, L.~Yuan, Z.~Liu, L.~Zhang, and N.~Vasconcelos, ``Micronet: Improving image recognition with extremely low flops,'' {\em 2021 IEEE/CVF International Conference on Computer Vision (ICCV)}, pp.~458--467, 2021.

\bibitem{lcnet}
W.~Xia, H.~Yin, X.~Dai, and N.~K. Jha, ``Fully dynamic inference with deep neural networks,'' {\em IEEE Transactions on Emerging Topics in Computing}, vol.~10, pp.~962--972, 2022.

\bibitem{drive_in_a_day}
A.~Kendall, J.~Hawke, D.~Janz, P.~Mazur, D.~Reda, J.~M. Allen, V.-D. Lam, A.~Bewley, and A.~Shah, ``Learning to drive in a day,'' {\em 2019 International Conference on Robotics and Automation (ICRA)}, pp.~8248--8254, 2019.

\bibitem{inhand_reorientation}
T.~Chen, J.~Xu, and P.~Agrawal, ``A system for general in-hand object re-orientation,'' in {\em Conference on Robot Learning}, pp.~297--307, PMLR, 2022.

\bibitem{learning_to_fly}
J.~Panerati, H.~Zheng, S.~Zhou, J.~Xu, A.~Prorok, A.~P. S.~U. of~Toronto Institute~for A~Studies, V.~I. for Artificial~Intelligence, and U.~of~Cambridge, ``Learning to fly—a gym environment with pybullet physics for reinforcement learning of multi-agent quadcopter control,'' {\em 2021 IEEE/RSJ International Conference on Intelligent Robots and Systems (IROS)}, pp.~7512--7519, 2021.

\bibitem{semantic_aware_uav_perception}
L.~Bartolomei, L.~Teixeira, and M.~Chli, ``Semantic-aware active perception for uavs using deep reinforcement learning,'' in {\em 2021 IEEE/RSJ International Conference on Intelligent Robots and Systems (IROS)}, pp.~3101--3108, 2021.

\bibitem{interp_e2e_driving}
J.~Chen, S.~E. Li, and M.~Tomizuka, ``Interpretable end-to-end urban autonomous driving with latent deep reinforcement learning,'' {\em arXiv preprint arXiv:2001.08726}, 2020.

\bibitem{airlearning}
S.~Krishnan, B.~Boroujerdian, W.~Fu, A.~Faust, and V.~J. Reddi, ``Air learning: a deep reinforcement learning gym for autonomous aerial robot visual navigation,'' {\em Mach. Learn.}, vol.~110, pp.~2501--2540, 2021.

\bibitem{glide}
Z.~Xie, X.~Da, B.~Babich, A.~Garg, and M.~van~de Panne, ``Glide: Generalizable quadrupedal locomotion in diverse environments with a centroidal model,'' {\em arXiv preprint arXiv:2104.09771}, 2021.

\bibitem{taxim}
Z.~Si and W.~Yuan, ``Taxim: An example-based simulation model for gelsight tactile sensors,'' {\em IEEE Robotics and Automation Letters}, vol.~7, no.~2, pp.~2361--2368, 2022.

\bibitem{walkminute}
N.~Rudin, D.~Hoeller, P.~Reist, and M.~Hutter, ``Learning to walk in minutes using massively parallel deep reinforcement learning,'' {\em ArXiv}, vol.~abs/2109.11978, 2021.

\bibitem{deepspeed_moe}
S.~Rajbhandari, C.~Li, Z.~Yao, M.~Zhang, R.~Y. Aminabadi, A.~A. Awan, J.~Rasley, and Y.~He, ``Deepspeed-moe: Advancing mixture-of-experts inference and training to power next-generation ai scale,'' in {\em ICML}, 2022.

\bibitem{hyperq}
``Nvidia inc, hyperq.'' \url{https://developer.download.nvidia.com/compute/DevZone/C/html_x64/6_Advanced/simpleHyperQ/doc/HyperQ.pdf}.
\newblock Accessed: 2023-07-21.

\bibitem{cudastream}
``Nvidia inc, cuda programming guide.'' \url{https://docs.nvidia.com/cuda/cuda-c-programming-guide/index.html#streams}.
\newblock Accessed: 2022-11-21.

\bibitem{cudagraph}
``Nvidia inc, getting started with cuda graphs.'' \url{https://developer.nvidia.com/blog/cuda-graphs/}.
\newblock Accessed: 2020-09-30.

\bibitem{atmi}
``Radeon open compute, atmi (asynchronous task and memory interface).'' \url{https://github.com/RadeonOpenCompute/atmi}.
\newblock Accessed: 2022-09-30.

\bibitem{task_superscalar}
Y.~Etsion, F.~Cabarcas, A.~Rico, A.~Ramirez, R.~M. Badia, E.~Ayguade, J.~Labarta, and M.~Valero, ``Task superscalar: An out-of-order task pipeline,'' in {\em 2010 43rd Annual IEEE/ACM International Symposium on Microarchitecture}, pp.~89--100, IEEE, 2010.

\bibitem{carbon}
S.~Kumar, C.~J. Hughes, and A.~D. Nguyen, ``Carbon: architectural support for fine-grained parallelism on chip multiprocessors,'' in {\em International Symposium on Computer Architecture}, 2007.

\bibitem{castillo_task_management}
E.~Castillo, L.~Alvarez, M.~Moret{\'o}, M.~Casas, E.~Vallejo, J.~L. Bosque, R.~Beivide, and M.~Valero, ``Architectural support for task dependence management with flexible software scheduling,'' {\em 2018 IEEE International Symposium on High Performance Computer Architecture (HPCA)}, pp.~283--295, 2018.

\bibitem{adm}
D.~S{\'a}nchez, R.~M. Yoo, and C.~E. Kozyrakis, ``Flexible architectural support for fine-grain scheduling,'' in {\em ASPLOS XV}, 2010.

\bibitem{oversubscribed_queues}
S.~Puthoor, X.~Tang, J.~Gross, and B.~M. Beckmann, ``Oversubscribed command queues in gpus,'' {\em Proceedings of the 11th Workshop on General Purpose GPUs}, 2018.

\bibitem{rl_data_generation}
J.~Gleeson, D.~Snider, Y.~Yang, M.~Gabel, E.~de~Lara, and G.~Pekhimenko, ``Optimizing data collection in deep reinforcement learning,'' {\em ArXiv}, vol.~abs/2207.07736, 2022.

\bibitem{rlscope}
J.~Gleeson, S.~Krishnan, M.~Gabel, V.~J. Reddi, E.~de~Lara, and G.~Pekhimenko, ``Rl-scope: Cross-stack profiling for deep reinforcement learning workloads,'' {\em ArXiv}, vol.~abs/2102.04285, 2021.

\bibitem{nasnet}
B.~Zoph, V.~Vasudevan, J.~Shlens, and Q.~V. Le, ``Learning transferable architectures for scalable image recognition,'' {\em 2018 IEEE/CVF Conference on Computer Vision and Pattern Recognition}, pp.~8697--8710, 2018.

\bibitem{amoebanet}
E.~Real, A.~Aggarwal, Y.~Huang, and Q.~V. Le, ``Regularized evolution for image classifier architecture search,'' in {\em AAAI}, 2019.

\bibitem{darts}
H.~Liu, K.~Simonyan, and Y.~Yang, ``Darts: Differentiable architecture search,'' {\em arXiv preprint arXiv:1806.09055}, 2018.

\bibitem{randomwire}
S.~Xie, A.~Kirillov, R.~B. Girshick, and K.~He, ``Exploring randomly wired neural networks for image recognition,'' {\em 2019 IEEE/CVF International Conference on Computer Vision (ICCV)}, pp.~1284--1293, 2019.

\bibitem{nasood}
H.~Bai, F.~Zhou, L.~Hong, N.~Ye, S.-H.~G. Chan, and Z.~Li, ``Nas-ood: Neural architecture search for out-of-distribution generalization,'' {\em 2021 IEEE/CVF International Conference on Computer Vision (ICCV)}, pp.~8300--8309, 2021.

\bibitem{condconv}
B.~Yang, G.~Bender, Q.~V. Le, and J.~Ngiam, ``Condconv: Conditionally parameterized convolutions for efficient inference,'' in {\em NeurIPS}, 2019.

\bibitem{speechmoe}
Z.~You, S.~Feng, D.~Su, and D.~Yu, ``Speechmoe: Scaling to large acoustic models with dynamic routing mixture of experts,'' {\em arXiv preprint arXiv:2105.03036}, 2021.

\bibitem{outrageously_large_moe}
N.~M. Shazeer, A.~Mirhoseini, K.~Maziarz, A.~Davis, Q.~V. Le, G.~E. Hinton, and J.~Dean, ``Outrageously large neural networks: The sparsely-gated mixture-of-experts layer,'' {\em ArXiv}, vol.~abs/1701.06538, 2017.

\bibitem{mobilenet}
M.~Sandler, A.~G. Howard, M.~Zhu, A.~Zhmoginov, and L.-C. Chen, ``Mobilenetv2: Inverted residuals and linear bottlenecks,'' {\em 2018 IEEE/CVF Conference on Computer Vision and Pattern Recognition}, pp.~4510--4520, 2018.

\bibitem{ios}
Y.~Ding, L.~Zhu, Z.~Jia, G.~Pekhimenko, and S.~Han, ``Ios: Inter-operator scheduler for cnn acceleration,'' {\em ArXiv}, vol.~abs/2011.01302, 2021.

\bibitem{nimble}
W.~Kwon, G.-I. Yu, E.~Jeong, and B.-G. Chun, ``Nimble: Lightweight and parallel gpu task scheduling for deep learning,'' in {\em NeurIPS}, 2020.

\bibitem{ooobackprop}
H.~Oh, J.~Lee, H.~Kim, and J.~Seo, ``Out-of-order backprop: an effective scheduling technique for deep learning,'' {\em Proceedings of the Seventeenth European Conference on Computer Systems}, 2022.

\bibitem{tensorflow}
M.~Abadi, P.~Barham, J.~Chen, Z.~Chen, A.~Davis, J.~Dean, M.~Devin, S.~Ghemawat, G.~Irving, M.~Isard, M.~Kudlur, J.~Levenberg, R.~Monga, S.~Moore, D.~G. Murray, B.~Steiner, P.~A. Tucker, V.~Vasudevan, P.~Warden, M.~Wicke, Y.~Yu, and X.~Zhang, ``Tensorflow: A system for large-scale machine learning,'' {\em ArXiv}, vol.~abs/1605.08695, 2016.

\bibitem{pytorch}
A.~Paszke, S.~Gross, F.~Massa, A.~Lerer, J.~Bradbury, G.~Chanan, T.~Killeen, Z.~Lin, N.~Gimelshein, L.~Antiga, A.~Desmaison, A.~K{\"o}pf, E.~Yang, Z.~DeVito, M.~Raison, A.~Tejani, S.~Chilamkurthy, B.~Steiner, L.~Fang, J.~Bai, and S.~Chintala, ``Pytorch: An imperative style, high-performance deep learning library,'' {\em ArXiv}, vol.~abs/1912.01703, 2019.

\bibitem{mxnet}
T.~Chen, M.~Li, Y.~Li, M.~Lin, N.~Wang, M.~Wang, T.~Xiao, B.~Xu, C.~Zhang, and Z.~Zhang, ``Mxnet: A flexible and efficient machine learning library for heterogeneous distributed systems,'' {\em ArXiv}, vol.~abs/1512.01274, 2015.

\bibitem{hsa_barrier}
S.~Puthoor, A.~M. Aji, S.~Che, M.~Daga, W.~Wu, B.~M. Beckmann, and G.~P. Rodgers, ``Implementing directed acyclic graphs with the heterogeneous system architecture,'' {\em Proceedings of the 9th Annual Workshop on General Purpose Processing using Graphics Processing Unit}, 2016.

\bibitem{hsa}
{HSA Foundation}, ``Hsa standard,'' 2017.
\newblock \url{http://hsafoundation.com/standards/}, Last accessed on 2023-02-14.

\bibitem{atos}
Y.~Chen, B.~Brock, S.~D. Porumbescu, A.~Bulucc, K.~A. Yelick, and J.~D. Owens, ``Atos: A task-parallel gpu dynamic scheduling framework for dynamic irregular computations,'' {\em ArXiv}, vol.~abs/2112.00132, 2021.

\bibitem{cdp}
``Nvidia inc, cuda dynamic parallelism.'' \url{https://developer.nvidia.com/blog/cuda-dynamic-parallelism-api-principles/}.
\newblock Accessed: 2022-09-30.

\bibitem{device_enqueue}
``Amd inc, rocm device enqueue.'' \url{https://sep5.readthedocs.io/en/latest/Programming_Guides/Opencl-programming-guide.html#device-side-enqueue}.
\newblock Accessed: 2022-09-30.

\bibitem{juggler}
M.~E. Belviranli, S.~Lee, J.~S. Vetter, and L.~N. Bhuyan, ``Juggler: a dependence-aware task-based execution framework for gpus,'' {\em Proceedings of the 23rd ACM SIGPLAN Symposium on Principles and Practice of Parallel Programming}, 2018.

\bibitem{whippletree}
M.~Steinberger, M.~Kenzel, P.~Boechat, B.~Kerbl, M.~Dokter, and D.~Schmalstieg, ``Whippletree: task-based scheduling of dynamic workloads on the gpu,'' {\em ACM Trans. Graph.}, vol.~33, pp.~228:1--228:11, 2014.

\bibitem{gpuocelot}
N.~Farooqui, A.~Kerr, G.~Diamos, S.~Yalamanchili, and K.~Schwan, ``A framework for dynamically instrumenting gpu compute applications within gpu ocelot,'' in {\em Proceedings of the Fourth Workshop on General Purpose Processing on Graphics Processing Units}, pp.~1--9, 2011.

\bibitem{accelsim}
M.~Khairy, Z.~Shen, T.~M. Aamodt, and T.~G. Rogers, ``Accel-sim: An extensible simulation framework for validated gpu modeling,'' {\em 2020 ACM/IEEE 47th Annual International Symposium on Computer Architecture (ISCA)}, pp.~473--486, 2020.

\bibitem{mujoco}
E.~Todorov, T.~Erez, and Y.~Tassa, ``Mujoco: A physics engine for model-based control,'' in {\em 2012 IEEE/RSJ international conference on intelligent robots and systems}, pp.~5026--5033, IEEE, 2012.

\bibitem{cityscape}
M.~Cordts, M.~Omran, S.~Ramos, T.~Rehfeld, M.~Enzweiler, R.~Benenson, U.~Franke, S.~Roth, and B.~Schiele, ``The cityscapes dataset for semantic urban scene understanding,'' in {\em Proceedings of the IEEE conference on computer vision and pattern recognition}, pp.~3213--3223, 2016.

\bibitem{efficientnet}
M.~Tan and Q.~Le, ``Efficientnet: Rethinking model scaling for convolutional neural networks,'' in {\em International conference on machine learning}, pp.~6105--6114, PMLR, 2019.

\bibitem{squeezenet}
F.~N. Iandola, M.~W. Moskewicz, K.~Ashraf, S.~Han, W.~J. Dally, and K.~Keutzer, ``Squeezenet: Alexnet-level accuracy with 50x fewer parameters and <1mb model size,'' {\em ArXiv}, vol.~abs/1602.07360, 2016.

\bibitem{pt_study}
K.~Gupta, J.~A. Stuart, and J.~D. Owens, ``A study of persistent threads style gpu programming for gpgpu workloads,'' {\em 2012 Innovative Parallel Computing (InPar)}, pp.~1--14, 2012.

\bibitem{pt_raytraversal}
T.~Aila and S.~Laine, ``Understanding the efficiency of ray traversal on gpus,'' {\em Proceedings of the Conference on High Performance Graphics 2009}, 2009.

\bibitem{daydream}
H.~Zhu, A.~Phanishayee, and G.~Pekhimenko, ``Daydream: Accurately estimating the efficacy of optimizations for {DNN} training,'' in {\em 2020 USENIX Annual Technical Conference (USENIX ATC 20)}, pp.~337--352, USENIX Association, July 2020.

\bibitem{onetbb}
J.~Reinders, M.~J. Voss, P.~Reble, and R.~Asenjo-Plaza, ``++ for heterogeneous programming: oneapi (dpc++ and onetbb),'' in {\em C++ for Heterogeneous Programming: oneAPI (DPC++ and oneTBB)}, 2020.

\bibitem{cilk}
R.~D. Blumofe, C.~F. Joerg, B.~C. Kuszmaul, C.~E. Leiserson, K.~H. Randall, and Y.~Zhou, ``Cilk: an efficient multithreaded runtime system,'' in {\em PPOPP '95}, 1995.

\bibitem{openmp}
L.~Dagum and R.~Menon, ``Openmp: an industry standard api for shared-memory programming,'' in {\em OpenMP: an industry standard API for shared-memory programming}, 1998.

\bibitem{sarc}
A.~Ram{\'i}rez, F.~Cabarcas, B.~H.~H. Juurlink, M.~Alvarez-Mesa, F.~S{\'a}nchez, A.~Azevedo, C.~Meenderinck, C.~B. Ciobanu, S.~Isaza, and G.~Gaydadjiev, ``The sarc architecture,'' {\em IEEE Micro}, vol.~30, pp.~16--29, 2010.

\bibitem{denovo}
B.~Choi, R.~Komuravelli, H.~Sung, R.~Smolinski, N.~Honarmand, S.~V. Adve, V.~S. Adve, N.~P. Carter, and C.-T. Chou, ``Denovo: Rethinking the memory hierarchy for disciplined parallelism,'' {\em 2011 International Conference on Parallel Architectures and Compilation Techniques}, pp.~155--166, 2011.

\bibitem{starss}
J.~Planas, R.~M. Badia, E.~Ayguad{\'e}, and J.~Labarta, ``Hierarchical task-based programming with starss,'' {\em The International Journal of High Performance Computing Applications}, vol.~23, pp.~284 -- 299, 2009.

\bibitem{openstream}
A.~Pop and A.~Cohen, ``Openstream: Expressiveness and data-flow compilation of openmp streaming programs,'' {\em ACM Trans. Archit. Code Optim.}, vol.~9, pp.~53:1--53:25, 2012.

\bibitem{multicoredataflow}
G.~Gupta and G.~S. Sohi, ``Dataflow execution of sequential imperative programs on multicore architectures,'' in {\em Proceedings of the 44th annual IEEE/ACM international symposium on Microarchitecture}, pp.~59--70, 2011.

\bibitem{serialization_sets}
M.~D. Allen, S.~Sridharan, and G.~S. Sohi, ``Serialization sets: a dynamic dependence-based parallel execution model,'' in {\em Proceedings of the 14th ACM SIGPLAN symposium on Principles and practice of parallel programming}, pp.~85--96, 2009.

\bibitem{dagee}
{AMD Research}, ``Dagee,'' 2017.
\newblock \url{https://github.com/AMDResearch/DAGEE.git}, Last accessed on 2023-02-14.

\bibitem{hipgraph}
{AMD Research}, ``Hipgraph,'' 2017.
\newblock \url{https://github.com/HipGraph/}, Last accessed on 2023-02-14.

\bibitem{ata}
A.~E. Helal, A.~M. Aji, M.~L. Chu, B.~M. Beckmann, and W.~chun Feng, ``Adaptive task aggregation for high-performance sparse solvers on gpus,'' {\em 2019 28th International Conference on Parallel Architectures and Compilation Techniques (PACT)}, pp.~324--336, 2019.

\bibitem{hyperplane_sweep_optimization}
A.~M. Kaushik, A.~M. Aji, M.~A. Hassaan, N.~Chalmers, N.~Wolfe, S.~Moe, S.~Puthoor, and B.~M. Beckmann, ``Optimizing hyperplane sweep operations using asynchronous multi-grain gpu tasks,'' {\em 2019 IEEE International Symposium on Workload Characterization (IISWC)}, pp.~59--69, 2019.

\bibitem{wireframe}
A.~Abdolrashidi, D.~Tripathy, M.~E. Belviranli, L.~N. Bhuyan, and D.~Wong, ``Wireframe: Supporting data-dependent parallelism through dependency graph execution in gpus,'' {\em 2017 50th Annual IEEE/ACM International Symposium on Microarchitecture (MICRO)}, pp.~600--611, 2017.

\bibitem{blockmaestro}
A.~Abdolrashidi, H.~A. Esfeden, A.~Jahanshahi, K.~Singh, N.~B. Abu-Ghazaleh, and D.~Wong, ``Blockmaestro: Enabling programmer-transparent task-based execution in gpu systems,'' {\em 2021 ACM/IEEE 48th Annual International Symposium on Computer Architecture (ISCA)}, pp.~333--346, 2021.

\bibitem{free_launch}
G.~Chen and X.~Shen, ``Free launch: Optimizing gpu dynamic kernel launches through thread reuse,'' {\em 2015 48th Annual IEEE/ACM International Symposium on Microarchitecture (MICRO)}, pp.~407--419, 2015.

\bibitem{klap}
I.~E. Hajj, J.~G{\'o}mez-Luna, C.~Li, L.-W. Chang, D.~S. Milojicic, and W.~mei W.~Hwu, ``Klap: Kernel launch aggregation and promotion for optimizing dynamic parallelism,'' {\em 2016 49th Annual IEEE/ACM International Symposium on Microarchitecture (MICRO)}, pp.~1--12, 2016.

\bibitem{dynamic_tb_launch}
J.~Wang, N.~Rubin, A.~Sidelnik, and S.~Yalamanchili, ``Dynamic thread block launch: A lightweight execution mechanism to support irregular applications on gpus,'' {\em 2015 ACM/IEEE 42nd Annual International Symposium on Computer Architecture (ISCA)}, pp.~528--540, 2015.

\bibitem{cortex}
P.~Fegade, T.~Chen, P.~Gibbons, and T.~Mowry, ``Cortex: A compiler for recursive deep learning models,'' {\em Proceedings of Machine Learning and Systems}, vol.~3, pp.~38--54, 2021.

\bibitem{janus}
E.~Jeong, S.~Cho, G.-I. Yu, J.~S. Jeong, D.-J. Shin, and B.-G. Chun, ``$\{$JANUS$\}$: fast and flexible deep learning via symbolic graph execution of imperative programs,'' in {\em 16th USENIX Symposium on Networked Systems Design and Implementation (NSDI 19)}, pp.~453--468, 2019.

\bibitem{cavs}
S.~Xu, H.~Zhang, G.~Neubig, W.~Dai, J.~K. Kim, Z.~Deng, Q.~Ho, G.~Yang, and E.~P. Xing, ``Cavs: An efficient runtime system for dynamic neural networks,'' in {\em 2018 USENIX Annual Technical Conference (USENIX ATC 18)}, pp.~937--950, 2018.

\bibitem{nimble_dynamicnn}
H.~Shen, J.~Roesch, Z.~Chen, W.~Chen, Y.~Wu, M.~Li, V.~Sharma, Z.~Tatlock, and Y.~Wang, ``Nimble: Efficiently compiling dynamic neural networks for model inference,'' {\em ArXiv}, vol.~abs/2006.03031, 2021.

\bibitem{deepframeworks_recursion}
E.~Jeong, J.~S. Jeong, S.~Kim, G.-I. Yu, and B.-G. Chun, ``Improving the expressiveness of deep learning frameworks with recursion,'' in {\em Proceedings of the Thirteenth EuroSys Conference}, pp.~1--13, 2018.

\bibitem{dl_dynamic_comp_basic}
M.~Looks, M.~Herreshoff, D.~S. Hutchins, and P.~Norvig, ``Deep learning with dynamic computation graphs,'' {\em ArXiv}, vol.~abs/1702.02181, 2017.

\bibitem{parameter_caching}
F.~Khorasani, H.~A. Esfeden, N.~B. Abu-Ghazaleh, and V.~Sarkar, ``In-register parameter caching for dynamic neural nets with virtual persistent processor specialization,'' {\em 2018 51st Annual IEEE/ACM International Symposium on Microarchitecture (MICRO)}, pp.~377--389, 2018.

\bibitem{ptask}
C.~J. Rossbach, J.~Currey, M.~Silberstein, B.~Ray, and E.~Witchel, ``Ptask: operating system abstractions to manage gpus as compute devices,'' in {\em Proceedings of the Twenty-Third ACM Symposium on Operating Systems Principles}, pp.~233--248, 2011.

\bibitem{dandelion}
C.~J. Rossbach, Y.~Yu, J.~Currey, J.-P. Martin, and D.~Fetterly, ``Dandelion: a compiler and runtime for heterogeneous systems,'' in {\em Proceedings of the Twenty-Fourth ACM Symposium on Operating Systems Principles}, pp.~49--68, 2013.

\bibitem{kernelet}
J.~Zhong and B.~He, ``Kernelet: High-throughput gpu kernel executions with dynamic slicing and scheduling,'' {\em IEEE Transactions on Parallel and Distributed Systems}, vol.~25, no.~6, pp.~1522--1532, 2013.

\bibitem{gpupool}
X.~Tan, {\em GPUPool: A Holistic Approach to Fine-Grained GPU Sharing in the Cloud}.
\newblock PhD thesis, University of Toronto (Canada), 2021.

\end{thebibliography}

% \clearpage
% \appendix
% \section{First appendix section}

\end{document}